\documentclass{aa}  
\usepackage{graphicx}
\usepackage{amsmath}
\usepackage{comment}

\newcommand{\bn}{\boldsymbol{n}}

\newcommand{\bO}{\boldsymbol{\omega}}
\newcommand{\osn}{\omega_{\text{s},\boldsymbol{n}}}
\newcommand{\osjml}{\omega_{\text{s},jml}}
\newcommand{\otn}{\theta_{\text{s},\boldsymbol{n}}}
\newcommand{\eexp}{\mathrm{e}}
\newcommand{\bJ}{\boldsymbol{J}}
\newcommand{\bth}{\boldsymbol{\theta}}

\newcommand{\Rb}{R_{\mathrm{b}}}

\newcommand{\Rsp}{R_{\mathrm{sp}}}
\newcommand{\Omegab}{\Omega_\mathrm{b}}
\newcommand{\Omp}{\Omega_\mathrm{p}}

\newcommand{\Omegasp}{\Omega_\mathrm{sp}}

\newcommand{\Kpc}{~\mathrm{kpc}}
 
\newcommand{\Gyr}{~\mathrm{Gyr}}

\newcommand{\kmsec}{~\mathrm{km}~\mathrm{s}^{-1}}

\newcommand{\de}{\mathrm{d}}

\newcommand{\Rg}{R_\mathrm{g}}

\newcommand{\RNum}[1]{\uppercase\expandafter{\romannumeral #1\relax}}

\newcommand{\tf}{t_\mathrm{f}}

\newcommand{\img}{\mathrm{i}}
\newcommand{\Rep}{\operatorname{Re}}

\newcommand{\hsp}{h_{\mathrm{sp}}}

\newcommand{\phis}{\varphi_{\mathrm{sp}}}
\newcommand{\phib}{\varphi_{\mathrm{b}}}
\newcommand{\alphab}{\alpha_{\mathrm{b}}}

\newcommand{\deriv}{\mathrm{d}}

\begin{document}

\title{Perturbed distribution functions with accurate action estimates for the Galactic disc}
\titlerunning{Perturbed distribution functions with accurate action estimates}
\authorrunning{Al Kazwini et al.}

\author{
H.~Al Kazwini \inst{1}
\and
Q.~Agobert \inst{1}
\and
A.~Siebert \inst{1}
\and
B.~Famaey \inst{1}
\and
G.~Monari \inst{1}
\and 
S.~Rozier \inst{1}
\and
P.~Ramos \inst{1}
\and
R.~Ibata \inst{1}
\and
S.~Gausland \inst{2}
\and 
C.~Rivi\`ere \inst{2}
\and 
D.~Spolyar \inst{3}
}

\institute{Universit\'e de Strasbourg, CNRS UMR 7550, Observatoire astronomique de Strasbourg, 11 rue de l'Universit\'e, 67000 Strasbourg, France
\and
Universit\'e de Strasbourg, UFR de Math\'ematique et d'Informatique, 7 rue Ren\'e Descartes, 67084 Strasbourg, France
\and
The Oskar Klein Centre for Cosmoparticle Physics, Department of Physics, Stockholm University, AlbaNova, 10691 Stockholm, Sweden
}

\date{Received; accepted}
 
  \abstract
   {In the Gaia era, understanding the effects of the perturbations of the Galactic disc is of major importance in the context of dynamical modelling. In this theoretical paper we extend previous work in which, making use of the epicyclic approximation, the linearized Boltzmann equation had been used to explicitly compute, away from resonances, the perturbed distribution function of a Galactic thin-disc population in the presence of a non-axisymmetric perturbation of constant amplitude. Here we improve this theoretical framework in two distinct ways in the new code that we present. First, we use better estimates for the action-angle variables away from quasi-circular orbits, computed from the {\tt AGAMA} software, and we present an efficient routine to numerically re-express any perturbing potential in these coordinates with a typical accuracy at the per cent level. The use of more accurate action estimates allows us to identify resonances such as the outer 1:1 bar resonance at higher azimuthal velocities than the outer Lindblad resonance (OLR), and to extend our previous theoretical results well above the Galactic plane, where we explicitly show how they differ from the epicyclic approximation. In particular, the displacement of resonances in velocity space as a function of height can in principle constrain the 3D structure of the Galactic potential. Second, we allow the perturbation to be time dependent, thereby allowing us to model the effect of transient spiral arms or a growing bar. The theoretical framework and tools presented here will be useful for a thorough analytical dynamical modelling of the complex velocity distribution of disc stars as measured by past and upcoming Gaia data releases.}


   \maketitle
%

\section{Introduction}

The natural canonical coordinate system of phase-space for Galactic dynamics and perturbation theory is the system of action-angle variables \citep{BT08}. In an axisymmetric system in equilibrium, the Jeans theorem implies that the phase-space distribution function (DF) of any stellar (or dark matter) component can be expressed solely as a function of the actions that are labelling the actual orbits \citep[e.g.][]{BinneyPiffl2015, ColeBinney2017}. However, the effect of various perturbers of the potential (e.g. the Galactic bar and spiral arms) must be included in this process, together with the response of the distribution function. Within the resonant regions, to fully capture the behaviour of the DF, one needs to  construct  for each perturber new orbital tori,  complete  with  a  new system  of  action-angle  variables \citep[e.g.][]{Monari2017, Binney2020a, Binney2020b}. Away from resonances, however, one can simply use the linearized Boltzmann equation. This also allows us to accurately identify the location of resonances.

This is particularly important in the context of the interpretation of recent data from the Gaia mission \citep{GaiaDR2, GaiaEDR3}, which revealed in exquisite detail the fine structure of stellar action space \citep[e.g.][]{trick19a, Monari2019a, Monari2019b}. While the existence of moving groups of dynamical origin had been known for a long time in local velocity space around the Sun \citep[e.g.][]{dehnen98,famaey05}, Gaia revealed their structure in exquisite detail \citep{Ramos2018} and also provided  an estimate of their age distribution \citep{Laporte2020}, together with the shape of the global velocity field away from the Sun within the Galactic disc \citep{Gaia2018}. One additional major finding of Gaia is the existence of a local phase-spiral in vertical height versus  vertical velocity in the  solar neighbourhood \citep{Antoja2018}, which might be related to a vertical perturbation of the disc, for example by  the Sagittarius dwarf galaxy \citep[e.g.][]{Laporte2019,BinneySchonrich2018, BlandHawthorn2020}.

In previous theoretical work, \citet{Monari2016} (hereafter M16) explicitly computed the response of an axisymmetric DF in action space, representing a Milky Way thin-disc stellar population, to a quasi-stationary three-dimensional spiral potential, using the epicyclic approximation to model the actions, which is only a valid approximation for quasi-circular orbits in the thin disc.  It was notably shown that the first-order moments of the perturbed DF then give rise to non-zero radial and vertical bulk flows (breathing modes). However, to treat perturbations away from the mid-plane, which is particularly important in the Gaia context, one cannot make use of the epicyclic approximation to compute action-angle variables. Moreover, it is well known that spiral modes in simulations can be transient, remaining stationary for only a few rotations \citep{SellwoodCarlberg2014}, and the response of the disc should be different during the period of rising or declining amplitude. The same can be true for the bar, whose amplitude and pattern speed can also grow or vary with time \citep[e.g.][]{Schonrichbar, Hilmi2020}.

Here we improve on this previous modelling of M16 in two ways. First, we  use a better estimate than the epicyclic approximation for the action-angle variables, relying on the torus mapping method of \cite{McgillBinney1990} to convert from actions and angles to positions and velocities, and on the St\"ackel fudge \citep{Binney2012, SandersBinney2016} for the reverse transformation. This will allow us to extend previous results to eccentric orbits and orbits wandering well above the Galactic plane. The routines developed and presented in this paper will also be of fundamental importance to study the vertical perturbations of the Galactic disc in further works. Second, we  let the perturbation evolve with time, thereby allowing us to model the effect of a growing bar.

The paper is organized as follows. In Sect.~2, after a short reminder of the theoretical framework of perturbed DF, already given in detail in M16, we present the method used to expand in Fourier series the perturbing potential within the new action-angle coordinate system. Then a comparison with the results in the epicyclic case is made in Sect.~3. In Sect.~4 we explore the temporal treatment of the DF for an analytic growth of the amplitude of the perturber. Finally, we conclude and discuss the possible future applications of these theoretical tools in Sect.~5. The Appendix is dedicated to the presentation of the code.

        \section{Perturbing potential and perturbed DF} \label{EpiStaApprox}
        
        \subsection{Action-angle variables}
        
        The canonical phase-space action-angle variables ($\bJ$,$\bth$) of an integrable system are obtained from a canonical transformation implicitly using Hamilton's characteristic function as a type 2 generating function. The actions $\bJ$ are defined as new generalized momenta corresponding to a closed path integral of the velocities along their corresponding canonically conjugate position variable, namely $J_i = \oint v_i {\rm d}x_i / (2\pi)$. Since this does not depend on time, these actions are integrals of motion, and the Hamiltonian can be expressed purely as a function of these actions. 
        
        It turns out that Galactic potentials are close enough to integrable systems that actions can be estimated for them. For quasi-circular orbits close to the Galactic plane, with separable motion in the vertical and horizontal directions, one can locally approximate the radial and vertical motions of an orbit with harmonic motions, which is known as the epicyclic approximation. The radial and vertical actions then simply correspond to the radial and vertical energies divided by their respective (radial and vertical) epicyclic frequency. However, the epicyclic approximation is no longer valid  when considering orbits with higher eccentricity, or with a large vertical amplitude. More precise ways of determining the action and angle coordinates have been devised. They typically differ depending on whether one wishes to  transform angles and actions to positions and velocities or instead to  make the reverse transformation from positions and velocities to actions and angles. In the first  case a very efficient method is the torus mapping method first introduced by \cite{McgillBinney1990} (see also \citealt{BinneyMcmillan2016} for a recent overview), while in the second case a St\"ackel fudge is generally used \citep{Binney2012, SandersBinney2016}.

The general idea of  torus mapping is to first express the Hamiltonian in the action-angle coordinates $({\bJ}_T, {\bth}_T)$ of a toy potential, for which the transformation to positions and velocities is fully known analytically. The algorithm then searches for a type 2 generating function $G(\bth_T, \bJ)$ expressed as a Fourier series expansion on the toy angles $\bth_T$, for which the Fourier coefficients are such that the Hamiltonian remains constant at constant $\bJ$. This generating function fully defines the canonical transformation from actions and angles to positions and velocities. For the inverse transformation, an estimate based on separable potentials can be used. These potentials are known as St\"ackel potentials \citep[e.g.][]{FamaeyDejonghe2003}, for which each generalized momentum depends on its conjugated position through three isolating integrals of the motion. These potentials are expressed in spheroidal coordinates defined by a focal distance. For a St\"ackel potential, this focal distance of the coordinate system is related to the first and second derivatives of the potential. One can thus use the true potential at any configuration space point to compute the equivalent focal distance {as if}  the potential were of St\"ackel form, and from there compute the corresponding isolating integrals of the motion and the actions.  In the following we  make use of both types of transformations, namely the torus mapping to express the potential in action-angle coordinates and the St\"ackel fudge to represent distribution functions in velocity space at a given configuration space point.
        
        We now  let $H_0(\bJ)$ be the Hamiltonian of the axisymmetric and time-independent zeroth order gravitational potential $\Phi_0$ of the Galaxy. The equations of motion connecting actions $\bJ$ and the canonically conjugate angles $\bth$ are then simply
        \begin{equation}
                \dot{\bth} = \frac{\partial H_0}{\partial \bJ} = \bO (\bJ), \quad \dot{\bJ} = -\frac{\partial H_0}{\partial \bth} = 0,
        \end{equation} with $\bO$ being the fundamental orbital frequencies. Thus, for a given orbit, the actions $\bJ$ are constant in time, defining an orbital torus on which the angles $\bth$ evolve linearly with time, according to $ \bth(t) = \bth_0 +  \bO t $. The Jeans theorem then tells us that the phase-space distribution function (DF) of an axisymmetric potential $f = f_0 (\bJ)$ is in equilibrium. In other words, $f_0$ is a solution of the collisionless Boltzmann equation: 
        \begin{equation} \label{CBE}
                \frac{\mathrm{d}f}{\mathrm{d}t} = 0.
        \end{equation}
        
        \subsection{Perturbed distribution functions}
        
                    \begin{figure*}
                \includegraphics[scale=0.4]{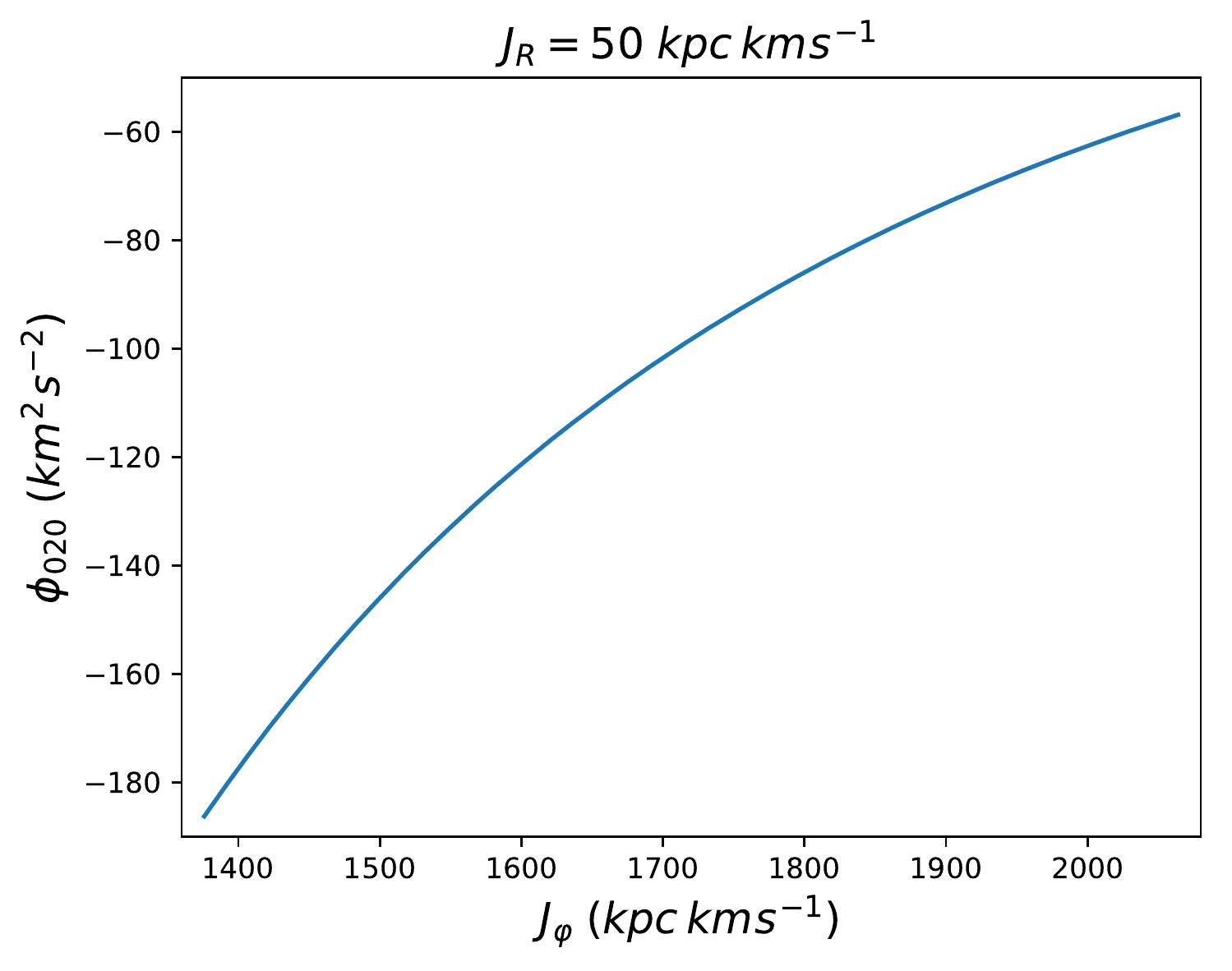}
                \includegraphics[scale=0.4]{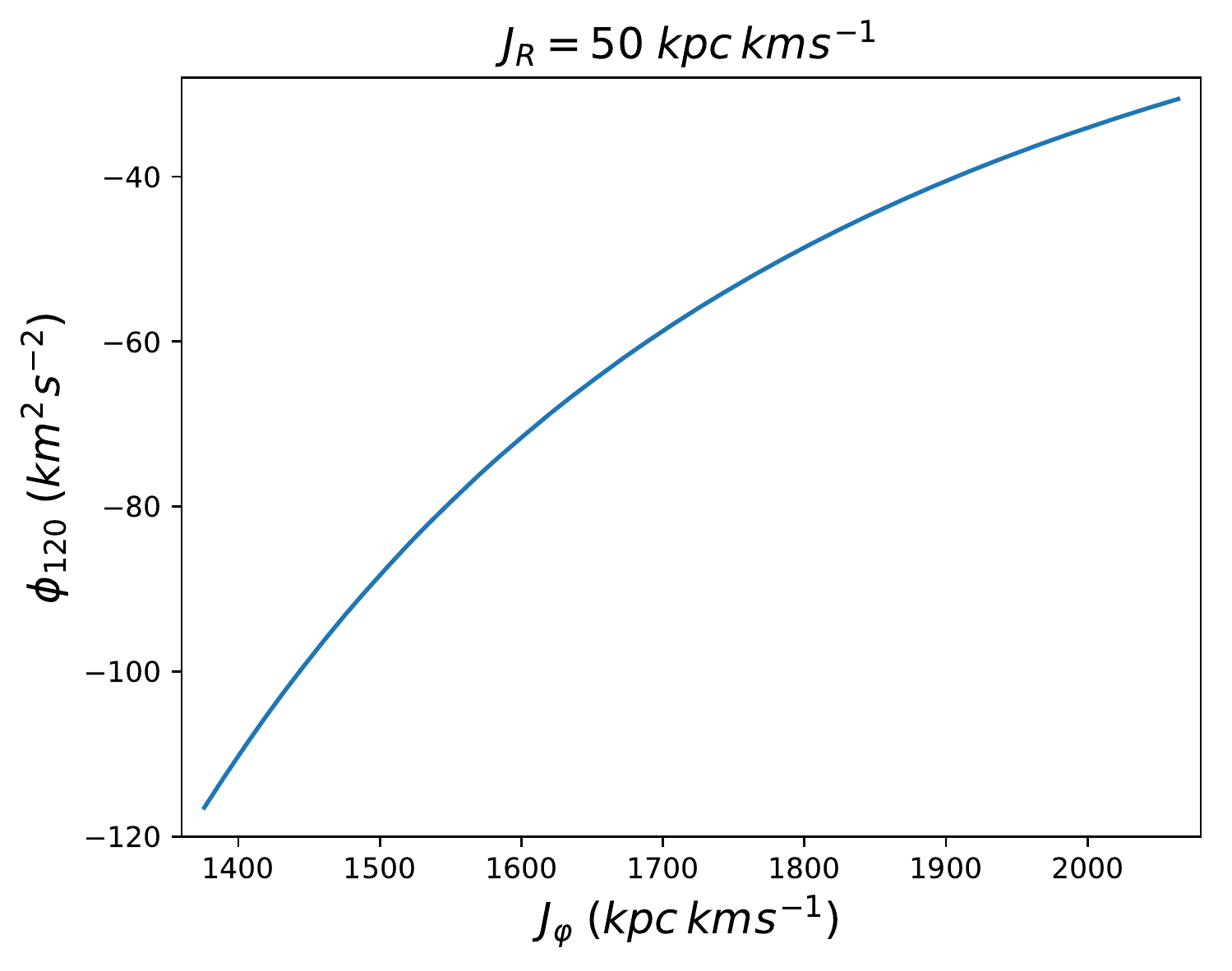}
                \includegraphics[scale=0.4]{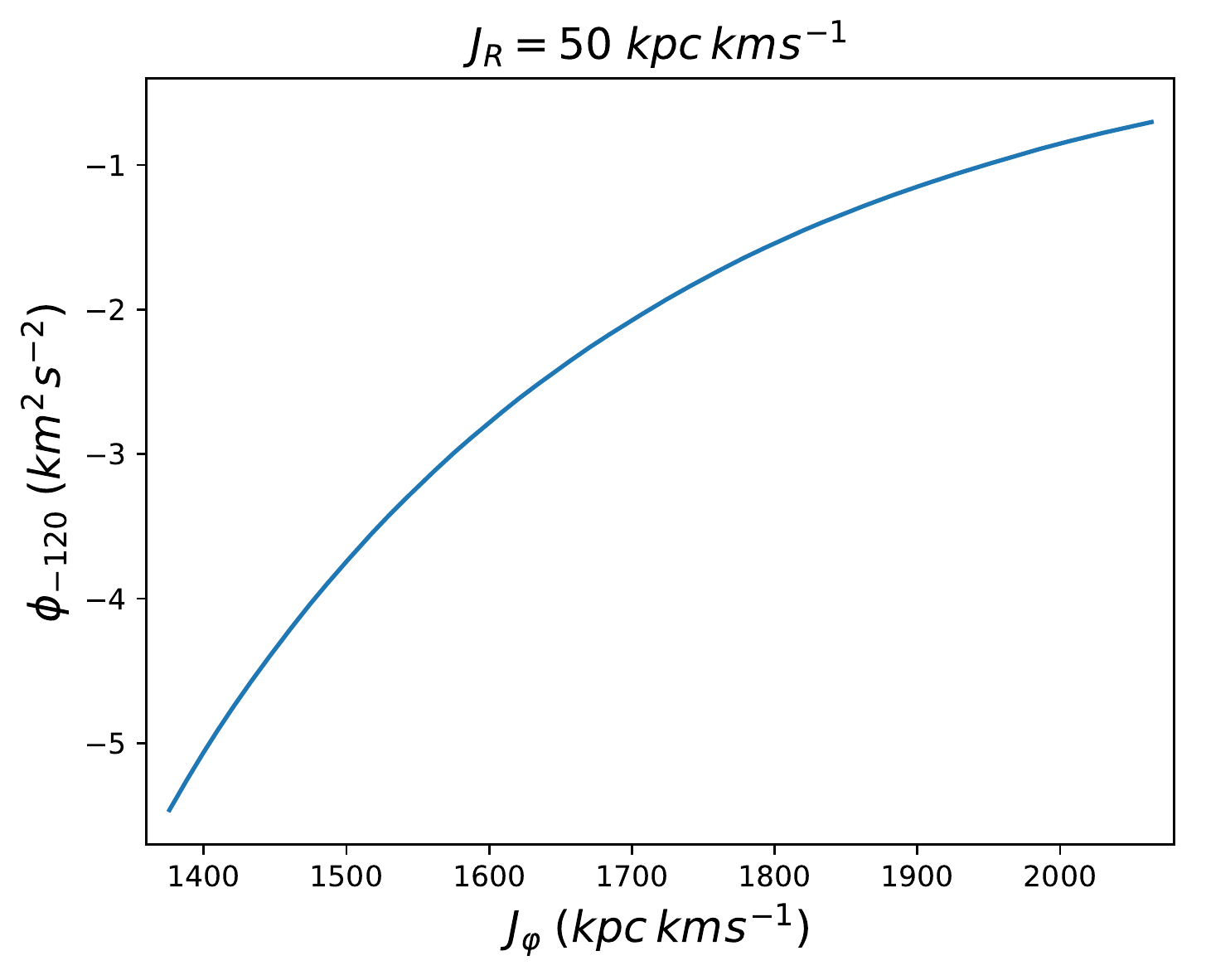}
                \includegraphics[scale=0.4]{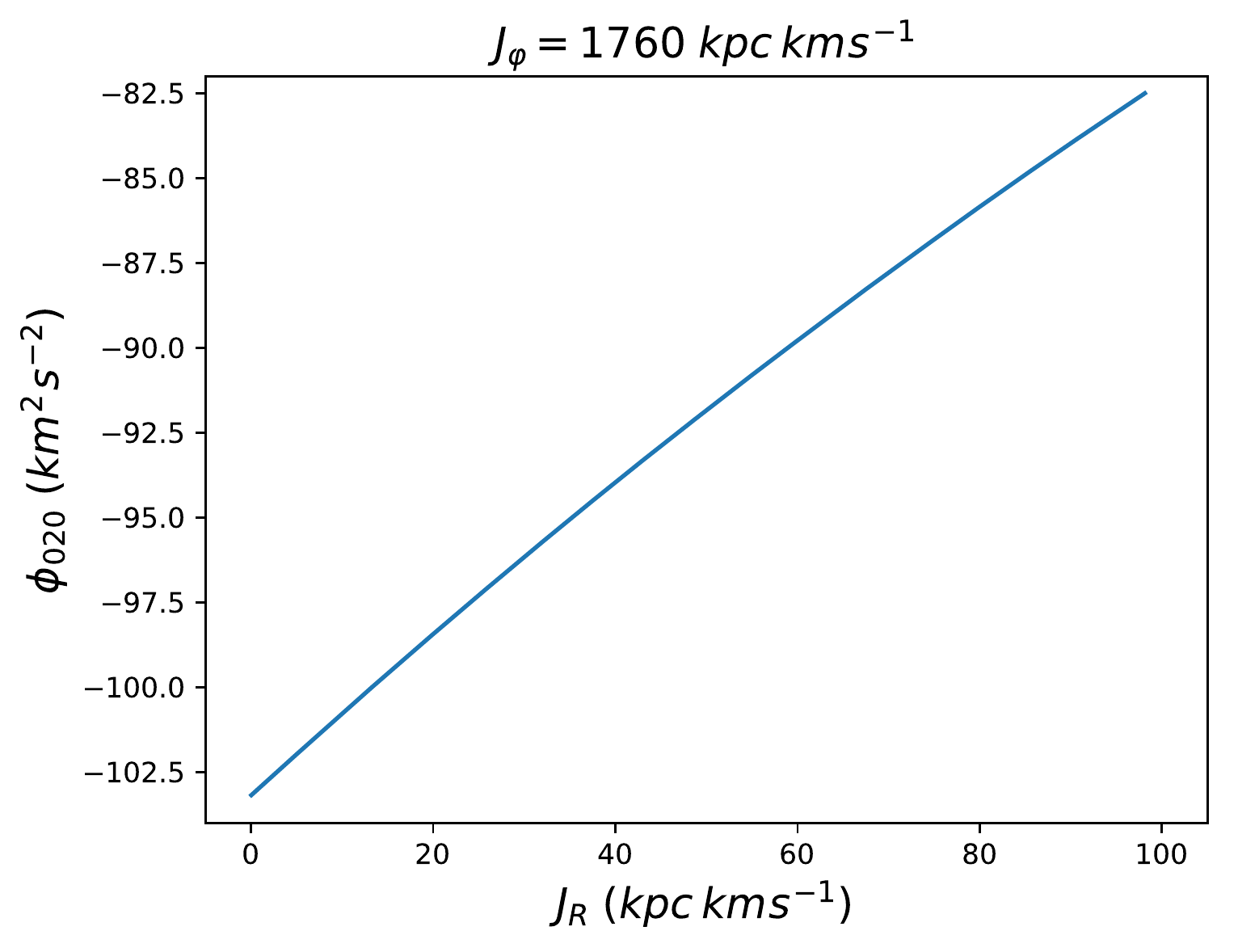}
                \includegraphics[scale=0.4]{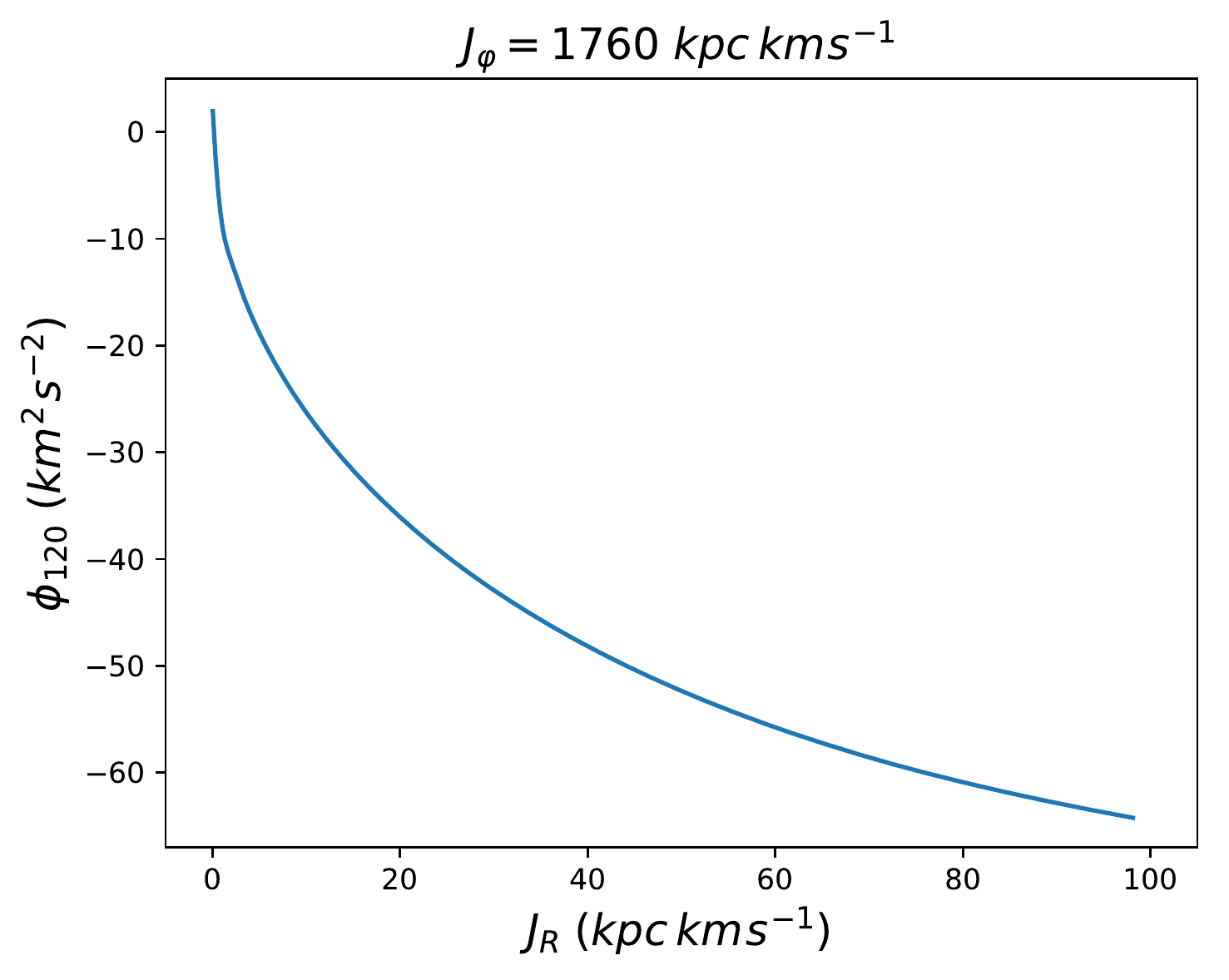}
                \includegraphics[scale=0.4]{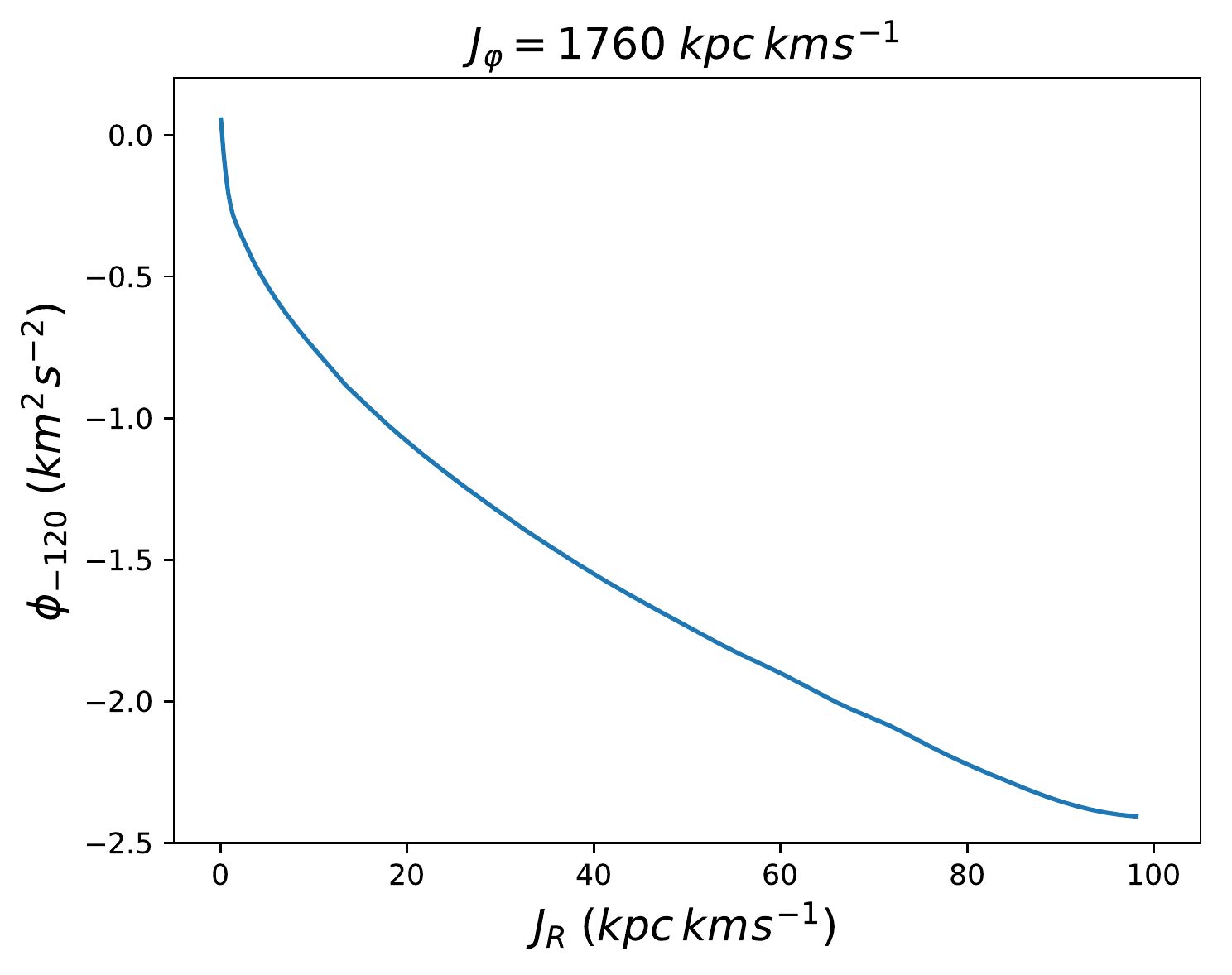}
                \caption{Variations in a few Fourier coefficients $\phi_{jml}(\bJ)$ of the bar potential from Sect.~2.4 as $J_R$ or $J_\varphi$ increase separately at $J_z=0$. The actions on the abscissa axis are in $\mathrm{kpc~km~s^{-1}}$. The curves are very smooth, which justifies our use of the cubic splines method to interpolate.}
                \label{CoeffBar}
        \end{figure*}   
        
                \begin{figure*}
            \centering     
                \includegraphics[scale=0.4]{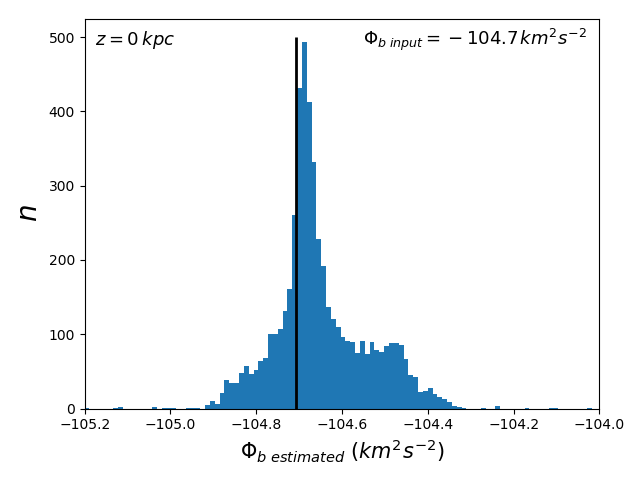}
                \includegraphics[scale=0.4]{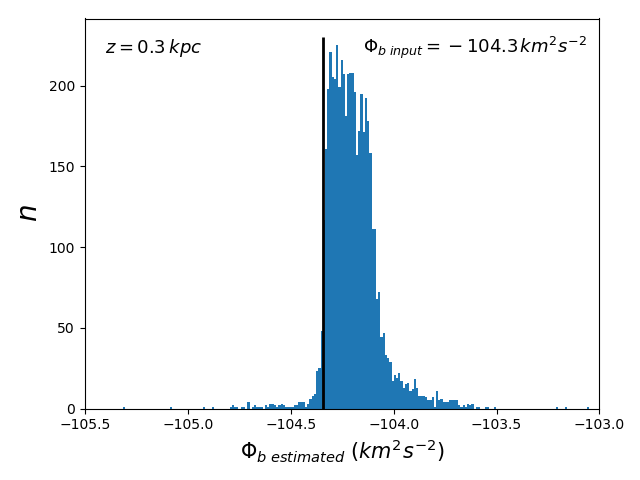}
                \includegraphics[scale=0.4]{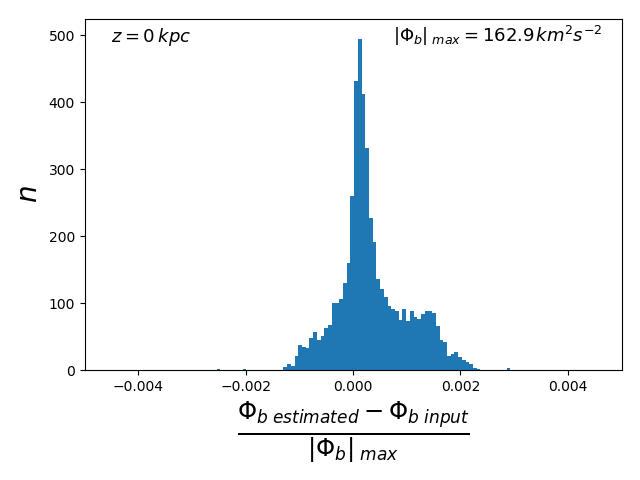}
                \includegraphics[scale=0.4]{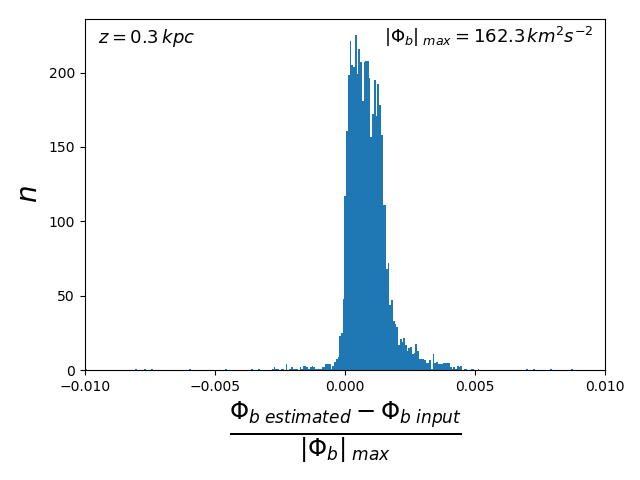}      
                \caption{Accuracy of the reconstruction of the bar potential . Left panel (top): Estimate of the bar potential from Sect.~2.4 at the Sun's position in the plane with 41 complex Fourier coefficients in Eq.~\ref{numFC} and the reconstruction using cubic splines. The vertical line denotes the true value. The value in the top right inset denotes the true value in physical units. The potential is always estimated at the same configuration space location (within the plane, at the Sun's position) but for different velocities. Left panel (bottom): Relative accuracy compared to the maximum value of the bar potential at the Sun's radius denoted in the top right inset. Right panel: Same, but  at $z = 0.3$~kpc with 231 complex Fourier coefficients. The typical accuracy is well below the per cent level, although with a slight bias towards lower amplitudes (i.e. $|\Phi_{\rm b \, estimated}| < |\Phi_{\rm b \, input}|$) above the plane.}
                \label{PotEstBar}
        \end{figure*}   
        
                 \begin{figure*}
            \centering
            \includegraphics[scale=0.4]{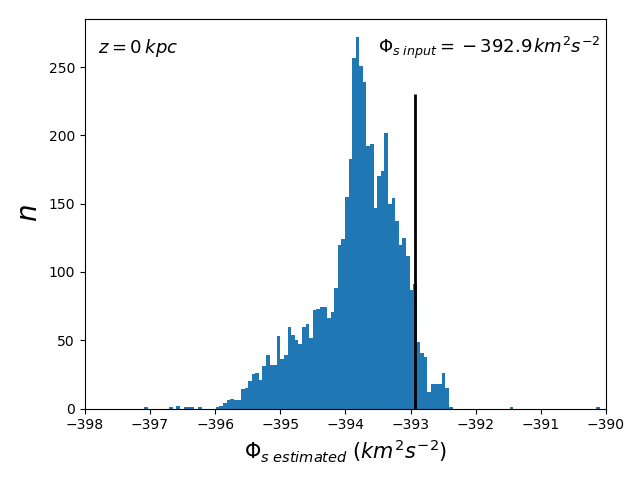}
            \includegraphics[scale=0.4]{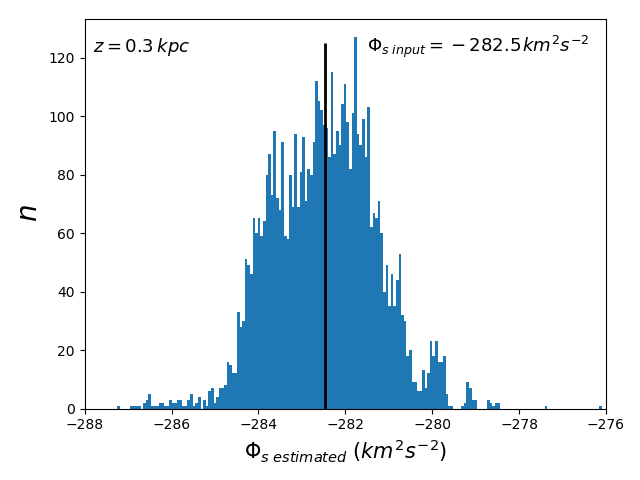}
                \includegraphics[scale=0.4]{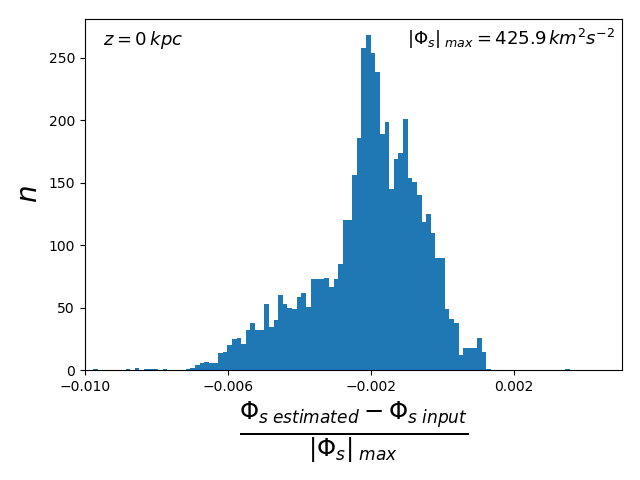}
                \includegraphics[scale=0.4]{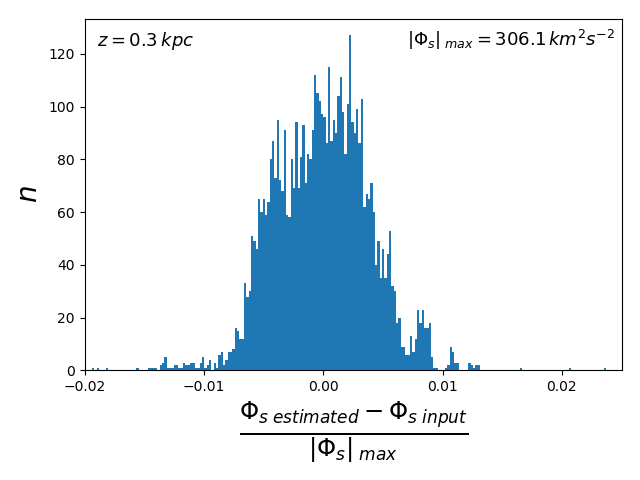}
                \caption{ Accuracy of the reconstruction of the spiral arms potential. Left panel (top): Estimate of the spiral arms potential from Sect.~2.5 at the Sun's position in the plane with 41 complex Fourier coefficients in Eq.~\ref{numFC} and the reconstruction using cubic splines. The vertical line denotes the true value. The value in the top right inset denotes the true value in physical units. The potential is always estimated at the same configuration space location (within the plane, at the Sun's position) but for different velocities. Left panel (bottom): Relative accuracy compared to the maximum value of the spiral arms potential at the Sun's radius denoted in the top right inset. Right panel: Same, but  at $z = 0.3$~kpc with 231 complex Fourier coefficients. The typical accuracy is again well below the per cent level in the plane, while it is around the per cent level above the plane.}
                \label{PotEstSpiral}
        \end{figure*}   
        
                \begin{figure*}
                \centering
                \includegraphics[scale=0.32]{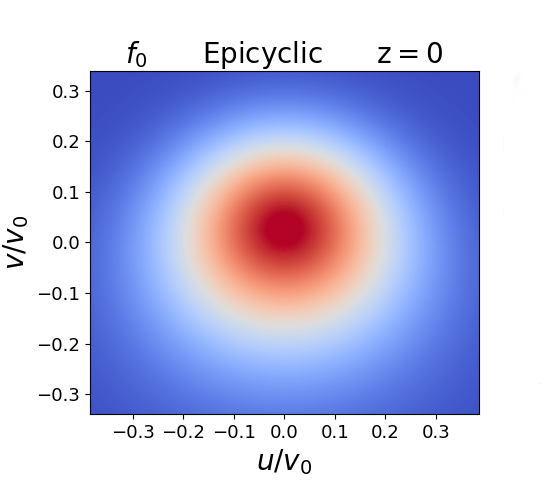}
                \includegraphics[scale=0.32]{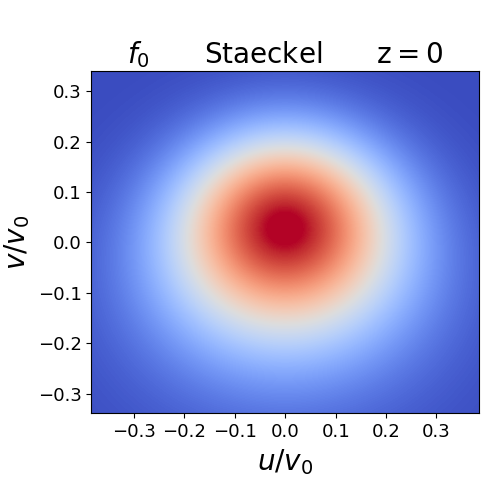}
                \includegraphics[scale=0.32]{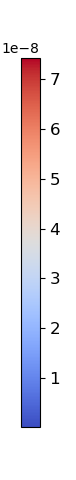}
                \includegraphics[scale=0.32]{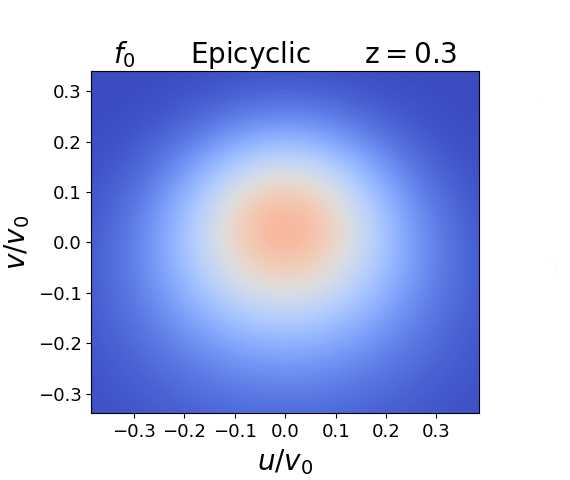}
                \includegraphics[scale=0.32]{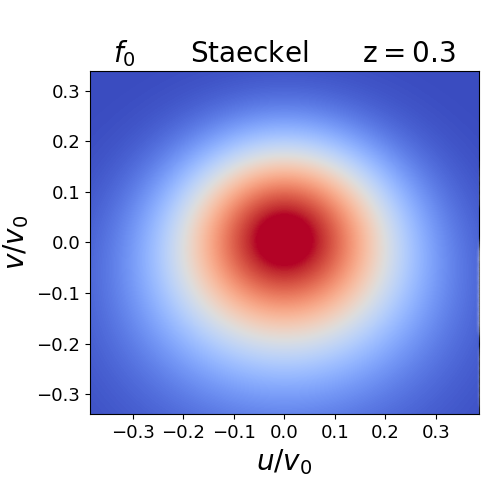}
                \includegraphics[scale=0.32]{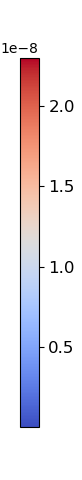}
                \caption{Local $uv$-plane stellar velocity distribution at axisymmetric equilibrium and for $w=0$, from Eq.~\ref{eq:f0} at ($R$, $z$, $\varphi$) = ($R_0$, $0$, $0$). Left panel: Epicyclic approximation in the plane (top) and at $z = 0.3$~kpc (bottom). Right panel: St\"ackel fudge with {\tt AGAMA} in the plane (top) and at $z = 0.3$~kpc (bottom).}
                \label{fig_f0}
        \end{figure*}

        We now let $\Phi_1$ be the potential of a small perturbation to the axisymmetric background potential $\Phi_0$ of the Galaxy, with an amplitude relative to this axisymmetric background $\epsilon \ll 1$. Then the total potential is $\Phi = \Phi_0 + \Phi_1$ and the DF becomes, to first order in $\epsilon$, $f = f_0 + f_1$, which is still a solution of the collisionless Boltzmann equation. Inserting $f = f_0 + f_1$ in Eq. (\ref{CBE}), and keeping only the terms of order $\epsilon$, leads to the \textit{linearized} collisionless Boltzmann equation 
        \begin{equation} \label{lCBE}
                \frac{\mathrm{d}f_1}{\mathrm{d}t} + [f_0,\Phi_1] = 0,
        \end{equation} where [,] is the Poisson bracket. Integrating Eq.~(\ref{lCBE}) over time, from $-\infty$ to the time $t$,  leads to 
        \begin{equation} \label{pert_df}
                f_1(\bJ,\bth,t) = \int_{-\infty}^{t} \mathrm{d}t' \frac{\partial f_0}{\partial \bJ'}(\bJ')\cdot\frac{\partial \Phi_1}{\partial \bth'}(\bJ',\bth',t'),
        \end{equation}
        where $\bJ'$ and $\bth'$ correspond to the actions and angles in the unperturbed case as a function of time (i.e. constant actions $\bJ'$ and angles evolving linearly).
        Since any function of the angles is $2\pi$-periodic in the angles, the perturbing potential $\Phi_1$ can be expanded in a Fourier series as 
        \begin{equation} \label{pert_pot}
                \Phi_1(\bJ,\bth,t) = \Rep \, \biggl\lbrace \sum_{\bn} \phi_{\bn}(\bJ, t) \,\eexp^{\img \bn \cdot \bth} \, \biggr\rbrace .
        \end{equation} 
        Hereafter we  consider in-plane perturbations such as spiral arms, meaning that we can write the time-varying Fourier coefficients in a non-rotating frame as $\phi_{\bn}(\bJ, t) = g(t) \, h(t) \, \phi_{\bn}(\bJ)$, where $g(t)$ controls the amplitude of the perturbation and $h(t)$ controls its pattern speed, with $h(t) = \eexp^{-\img m \Omp t}$, where $\Omp$ is the pattern speed of the perturbation and $m$ its azimuthal wave number (i.e. its multiplicity).  Hereafter we  mainly consider  $m=2$ perturbations. The vector index $\bn$ is a triplet of scalar integer indices $(j,k,l)$ running in principle from $-\infty$ to $\infty$, but in practice limited to a given range sufficient to approximate the perturbing potential. In the case of an $m$-fold in-plane perturbation, it is sufficient to take $k=m$. The main goal of this section is  to express typical non-axisymmetric perturbing potentials originally expressed in configuration space as such a Fourier series in action-angle space. The algorithm that we present in Sect.~2.3 can  be applied to any perturbing potential, however,  including non-plane symmetric vertical perturbations. 
        Once the potential is expressed as a function of angles and actions as in Eq.~(\ref{pert_pot}), then Eq.~(\ref{pert_df}) becomes
\begin{align} \label{pert_df_2}
        f_1(\bJ,\bth,t) = \Rep \, \biggl\lbrace \, &\img \frac{\partial f_0}{\partial \bJ}(\bJ) \cdot \sum_{\bn} \bn 
        \int_{-\infty}^{t} \mathrm{d}t' \phi_{\bn}(\bJ', t') \, \eexp^{\img\bn\cdot\bth'(t')} \biggr\rbrace.
\end{align}
In M16, assuming $\phi_{\bn}(\bJ', t') = g(t') \, h(t') \, \phi_{\bn}(\bJ)$, with $h(t') = \eexp^{-\img m \Omp t'}$, and the amplitude of the perturbing potential constant in time at present time ($g(t)=1$), and zero and constant in time at $-\infty$, led to the following explicit solution for $f_1(\bJ,\bth,t)$, 
\begin{equation}
\label{f1M16}
        f_1(\bJ,\bth,t) = \Rep \, \biggl\lbrace \frac{\partial f_0}{\partial \bJ}(\bJ) \cdot \sum_{\bn} \bn \phi_{\bn}(\bJ) \frac{\eexp^{\img\otn}}{\osn} \biggr\rbrace,
\end{equation}
where we defined
\begin{equation}
    \otn = \bn \cdot \bth - m \Omp t,
\end{equation}
\begin{equation}
    \osn = \bn \cdot \bO - m \Omp.
\label{eq:omegas}
\end{equation}
The subscript `s' stands for slow, because in the proximity of a resonance of the type $\osn = 0$, the angle $\otn$ evolves very slowly. One can also immediately see that the linearized solution above is valid only away from such resonances since it diverges for these orbits. Orbits near these resonances are actually trapped, and for them the determination of the linearly perturbed DF becomes inappropriate. Specific treatment for these resonant regions is required, which was addressed in  \citet{Monari2017} within the epicyclic approximation, and by \citet{Binney2020a} in a more general context.

Using the epicyclic approximation the Fourier coefficients of a spiral potential have been computed analytically in M16 with indices $\bn=(j,k,l)$ running over the values $j= \{-1, 0, 1\}$, $k=m=2$, and $l=\{-2, 0, 2\}$, and the perturbed distribution function away from resonances was then computed.
In the following we  extend the results of M16 to a more general estimate of the action-angle variables through the torus mapping method. The resulting DF is  plotted in velocity space by making use of the St\"ackel fudge. For both transformations we  use the  Action-based Galaxy
Modelling Architecture   \citep[{\tt AGAMA};][]{Vasiliev1, Vasiliev}.
        
        \subsection{Perturbing potential in actions and angles} \label{EpiApprox}

In previous work, M16 worked in the epicyclic approximation as it provides an analytical expression for evaluating actions and angles from cylindrical coordinates. The Fourier coefficients of the Fourier series expansion of the perturbing potentials were then also determined analytically within this approximation. Approximating the vertical component of the perturbing potential by a harmonic oscillator, the nine Fourier coefficients $\phi_{jml}$ were then limited to the range $j=\{-1,0,1\}$, corresponding to the $\theta_R$ oscillations of the potential, and $l=\{-2,0,2\}$, corresponding to the $\theta_z$ oscillations of the potential close to the Galactic plane.

However, the epicyclic approximation is only valid for nearly circular orbits, but not  when considering eccentric orbits. Hereafter the transformations from angles and actions to positions and velocities (and reciprocally) are  instead evaluated numerically with {\tt AGAMA} \citep[][]{Vasiliev}. The code makes use of  torus mapping to go from action-angle to position-velocity, and uses the St\"ackel fudge for the inverse transformation. Our goal now is to obtain the Fourier coefficients of a known perturbing potential using these numerically computed actions (instead of epicyclic). 

We proceed in the following way to evaluate Fourier coefficients of the perturbing potential in Eq.~(\ref{pert_pot}). The first step is to choose a set of actions within a range representing all the orbits of interest in the axisymmetric background configuration, each triplet of actions representing one of the orbits. For instance, in the solar neighbourhood we consider radial actions ranging from $0$ to $220 \, {\rm kpc} \, {\rm km} \, {\rm s}^{-1}$, azimuthal actions ranging from $1200$ to $2160 \, {\rm kpc} \, {\rm km} \, {\rm s}^{-1}$, and vertical actions ranging from $0$ to $26 \, {\rm kpc} \, {\rm km} \, {\rm s}^{-1}$ depending on the height above the Galactic plane. For each orbit, we then define an array of angles $(\theta_R,\theta_\varphi,\theta_z)$. These actions and angles can then all be converted to positions thanks to the torus machinery in {\tt AGAMA}. For each triplet of actions, a range of positions $(R,\varphi,z)$ is covered by the angles, and we look for the best-fitting coefficients $\phi_{jml}(J_R, J_z, J_{\varphi})$, satisfying the following equation (setting $t=0$ for the time being):
        \begin{align} \label{numFC}
                \Phi_1(R,\varphi,z) 
                &= \, \Rep \, \biggl\lbrace \sum_{j,l} \phi_{jml}(J_R, J_z, J_{\varphi}) \eexp^{\img(j\theta_R + m\theta_{\varphi} + l\theta_z)} \, \biggr\rbrace.
        \end{align} 
This is performed with the method of linear least squares using singular value decomposition, as proposed in chapter 15.4 of \citet{NumRec}. We then interpolate the value of the coefficient $\phi_{jml}$ with cubic splines, also  proposed in \citet{NumRec}, chapter 3.3. The number of Fourier coefficients is chosen to be high enough to ensure that all orbits passing through a given configuration space point yield the same value of the potential at this point within a relative accuracy of less than 1\%.

Concretely, we apply this hereafter to the potential of a central bar and of a two-armed spiral pattern. The background axisymmetric potential is chosen to be Model I from \citet{BT08}. This potential has a bulge described by a truncated oblate spheroidal power law; a gaseous disc with a hole at the centre; a stellar thin disc and a stellar thick disc, both with a scale-length of 2~kpc; and a dark halo with an oblate two-power-law profile. The galactocentric radius of the Sun is set at $R_0=8\,\Kpc$, and the local circular velocity is $v_0= 220 \, \kmsec$.

\subsection{Bar potential}

The potential we choose for the bar is a simple quadrupole potential \citep{Weinberg1994, Dehnen2000} with
        \begin{equation}
        \label{barpot}
                \Phi_{1,\mathrm{b}}(R,z,\varphi,t) = \Rep \, \biggl\lbrace \Phi_{\mathrm{a,b}}(R,z) \eexp^{\img m (\varphi-\phib-\Omegab t)} \biggr\rbrace,
        \end{equation} where $m=2$, $\Omegab$ is the pattern speed of the bar (expressed hereafter in multiples of the angular frequency at the Sun $\Omega_0=v_0/R_0$, where $v_0$ is the local circular velocity at the galactocentric radius of the Sun $R_0$), and the azimuth is defined with respect to a line   corresponding to the Galactic centre-Sun direction in the Milky Way, $\phib$ thus being the angle between the Sun and the long axis of the bar. We also choose
        \begin{equation}
        \label{barpot2}
                \Phi_{\mathrm{a,b}}(R,z) = -\alphab \dfrac{v_0^2}{3} \left(\dfrac{R_0}{\Rb}\right)^3 \left(\dfrac{R}{r}\right)^2 \left\lbrace
                \begin{array}{cc}
                        \left(\dfrac{r}{\Rb}\right)^{-3} & R \geq \Rb, \\
                        2-\left(\dfrac{r}{\Rb}\right)^{3} & R < \Rb,
                \end{array} \right.
        \end{equation} where $r^2 = R^2+z^2$ is the spherical radius, $\Rb$ is the length of the bar, and $\alphab$ represents the maximum ratio between the bar and axisymmetric background radial forces at the Sun’s galactocentric radius $R=R_0$.  We use hereafter, as a representative example, $\Rb = 0.625\,R_0$, $\phib = 25^{\mathrm{o}}$, and $\alphab = 0.01$. We  also consider two typical pattern speeds: $\Omegab = 1.89\,\Omega_0$ and $\Omegab = 1.16\,\Omega_0$.\\

    The bar potential is quite easy to reproduce using Fourier coefficients since it varies smoothly on orbits. Thus, for a study in the Galactic plane,  41 complex Fourier coefficients for each triplet of actions are sufficient to approximate the value of the potential with an accuracy  much better than 1\%. Here it should be noted that the potential of the bar oscillates along the azimuth at a given radius and that the relative accuracy can be ill-defined when the potential passes through zero. Therefore, we define here the relative accuracy with respect to the amplitude (i.e. the maximum value) of the bar potential at a given radius. The Fourier coefficients themselves vary smoothly, as illustrated in Fig.~\ref{CoeffBar}, which shows the variations of a few Fourier coefficients as $J_R$ and $J_{\varphi}$ increase separately, justifying the use of cubic-spline interpolation to get the value of the potential at a specific position. Figure~\ref{PotEstBar} demonstrates the accuracy of our method in reproducing the bar potential in the solar neighbourhood for different values of the local velocities. The potential is estimated at the same configuration space location for the whole range of relevant velocities, with a typical accuracy at the per cent level both in the plane and at $z=0.3$~kpc. The accuracy remains very good above the plane, although with a slight bias towards lower amplitudes than the true value. More complex Fourier coefficients are needed outside of the plane. This tool is of course not limited to any specific form of the perturbing potential, the only adjustable parameter being the number of Fourier coefficients necessary to recover a given  perturbing potential with a per cent-level accuracy.

\subsection{Spiral potential}

            \begin{figure*}
    \centering
                \includegraphics[scale=0.28]{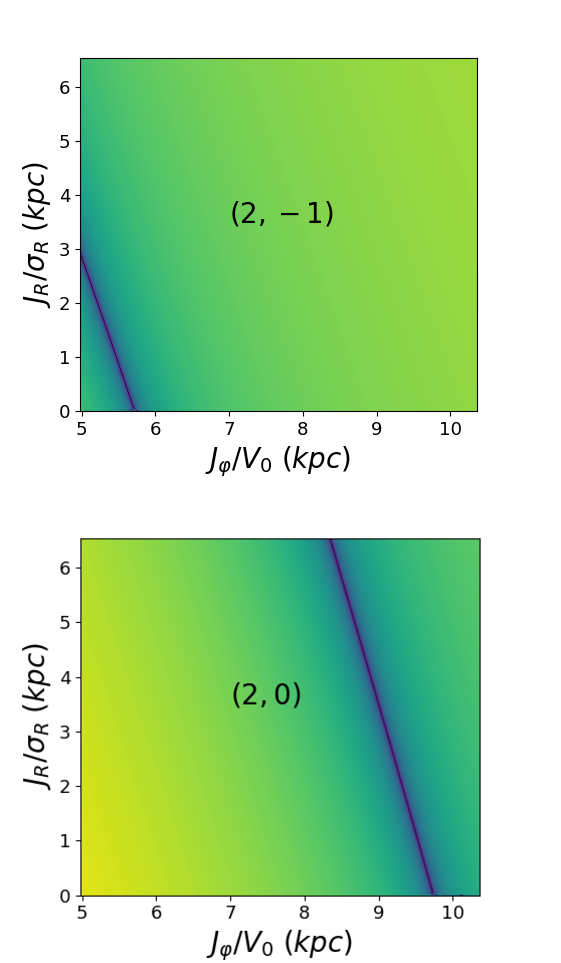}
                \includegraphics[scale=0.28]{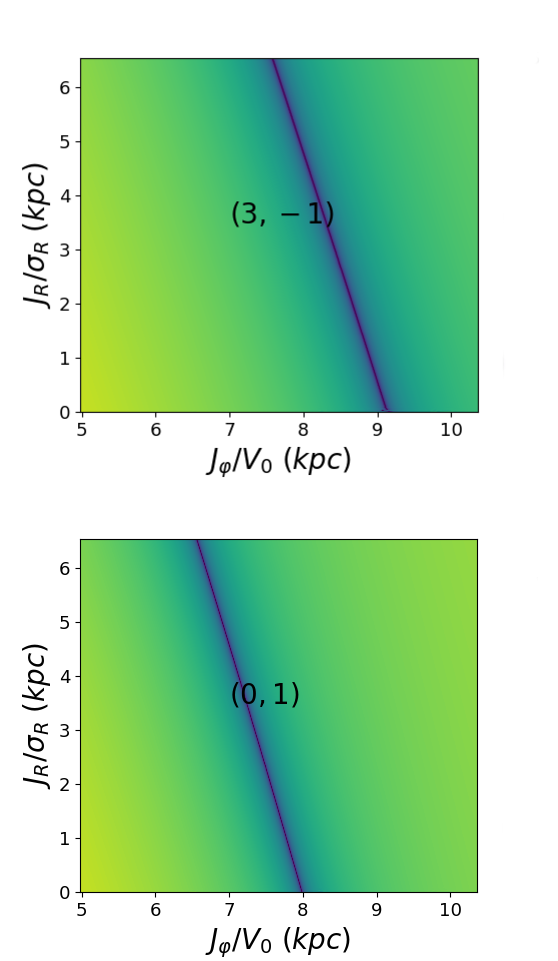}
                \includegraphics[scale=0.28]{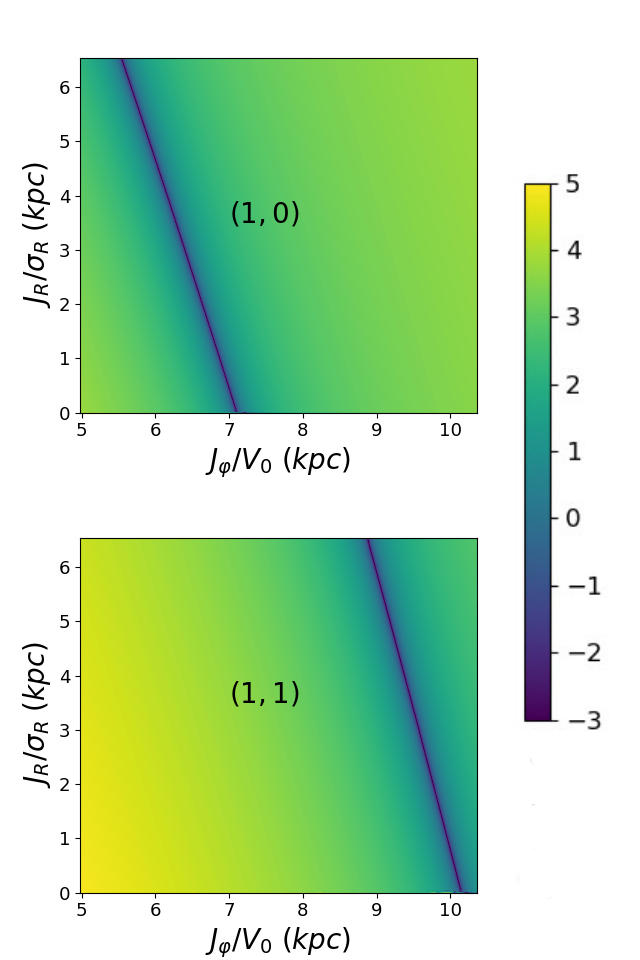}
                \caption{Values of ${\rm log}(\osjml)$ in the $(J_R,J_\varphi)$ plane with fixed $J_z = 10 \, \Kpc~\kmsec$, for a few  combinations of $(j,l)$ indices giving rise to resonant zones in action space (recalling that $m=2$). The pattern speed $\Omp$  here is $1.89 \, \Omega_0$. The two actions are renormalized by the radial velocity dispersion of the thin disc and the circular velocity at the Sun, respectively. The deep blue lines correspond to resonant zones. For instance, the $(1,0)$ case corresponds to the traditional OLR (for a non-zero $J_z$). Most other low-order combinations of indices did not give rise to any relevant resonant zone in the region of interest.}
                \label{resonanceBarJphi}
        \end{figure*}

    \begin{figure*}
        \centering
                \includegraphics[scale=0.28]{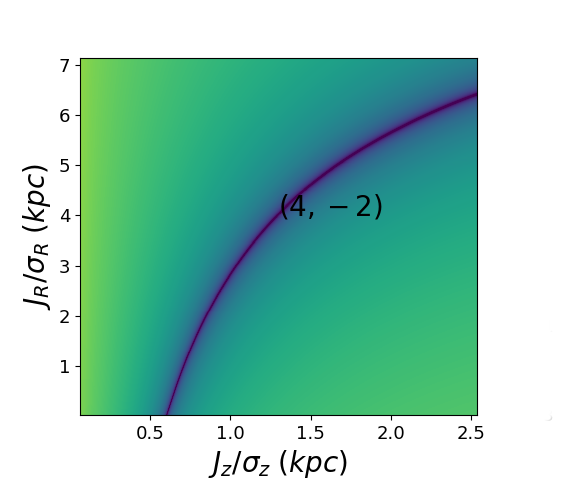}
                \includegraphics[scale=0.28]{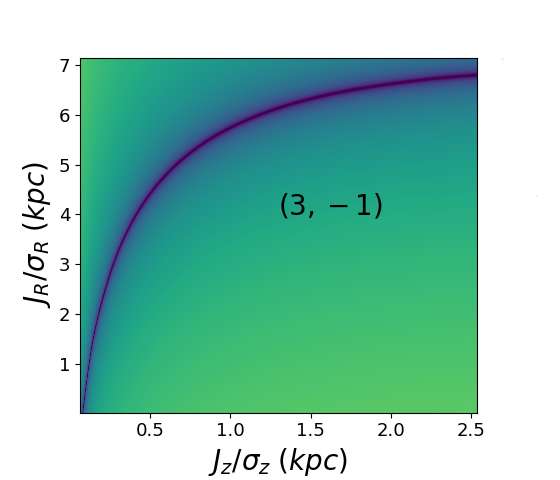}
                \includegraphics[scale=0.28]{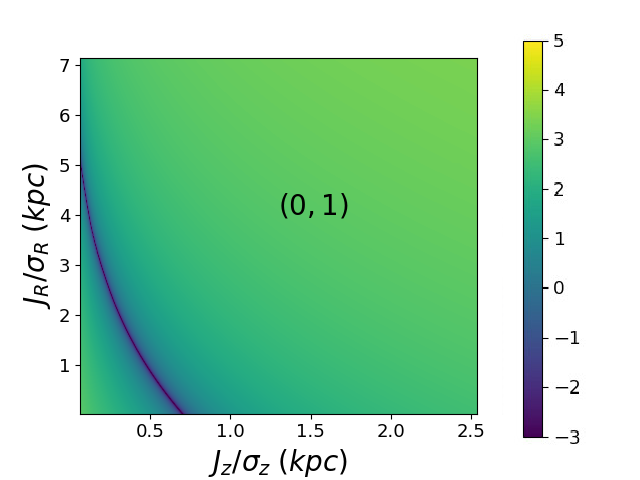}
                \caption{Values of ${\rm log}(\osjml)$ in the ($J_R$,$J_z$) plane with fixed $J_\varphi = 1759 \, \Kpc~\kmsec$ for different $(j,l)$ resonances. The pattern speed $\Omp$ is that of our fiducial central bar fixed at $1.89 \, \Omega_0$. The two actions are renormalized by the radial velocity dispersion and the vertical velocity dispersion of the thin disc at the Sun, respectively. The deep blue lines correspond to resonance zones. Most combinations of indices explored did not give rise to any relevant resonant zone in the region of interest.}
                \label{resonanceBarJz}
        \end{figure*}
        
            \begin{figure*}
        \centering
            \includegraphics[scale=0.28]{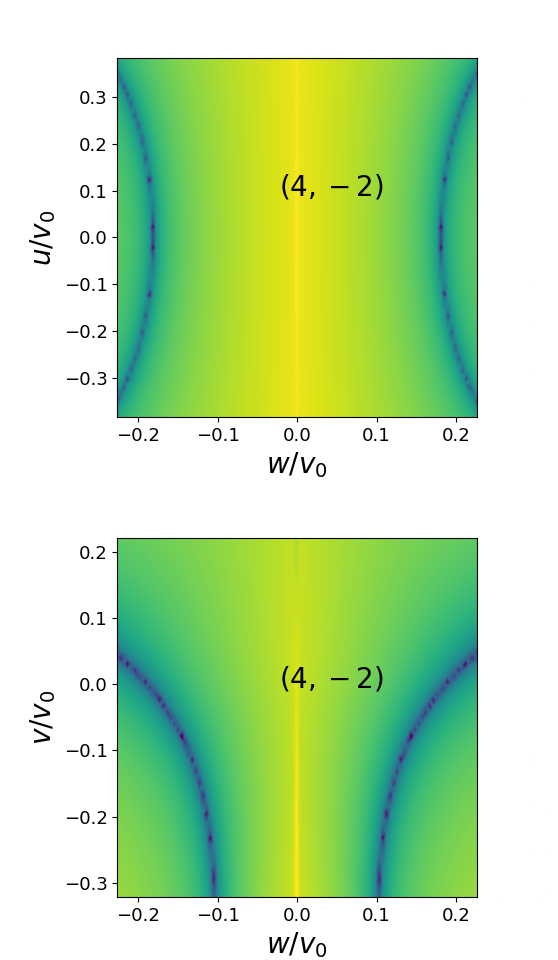}
            \includegraphics[scale=0.28]{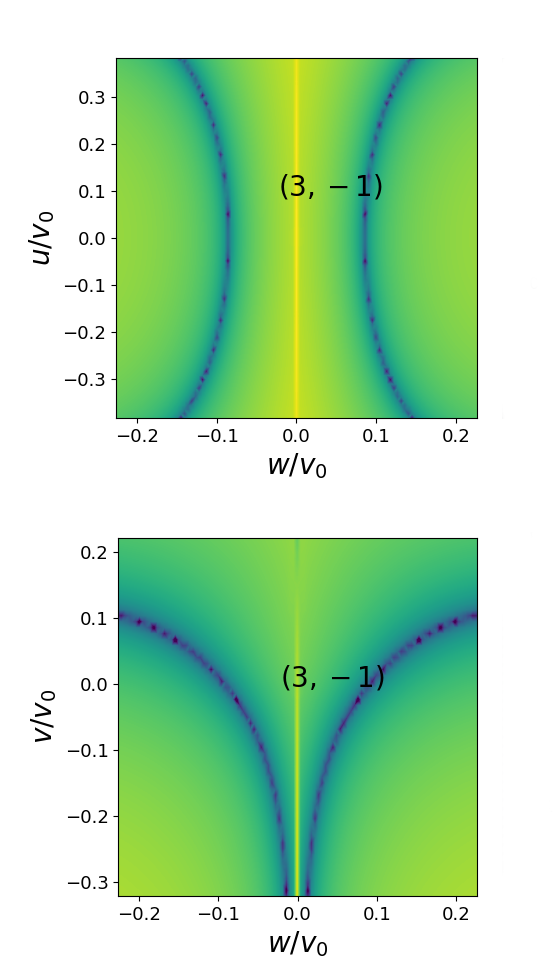}
            \includegraphics[scale=0.28]{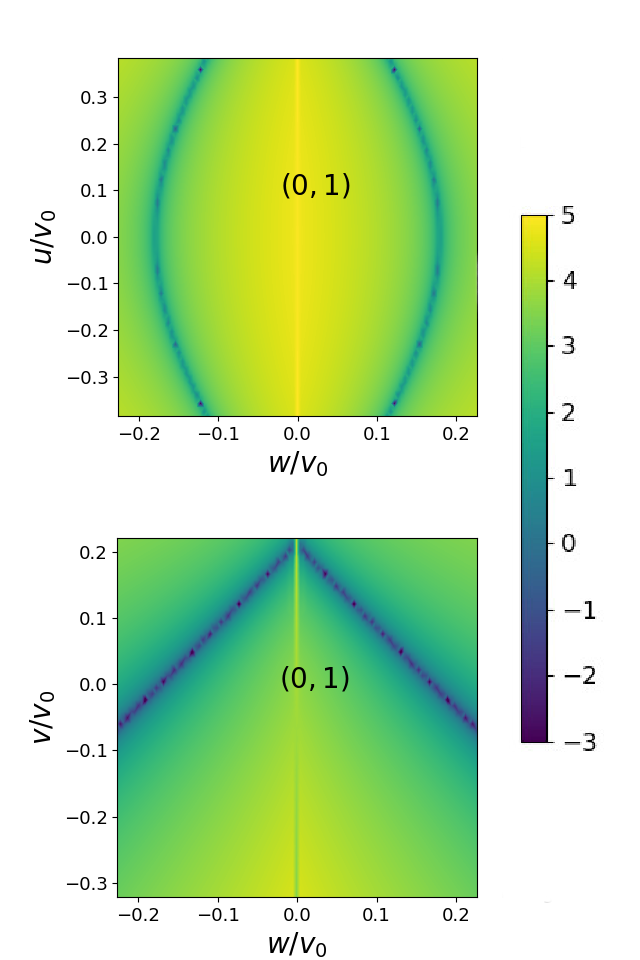}
            \caption{Values of $log( \osjml)$ in the $uw$-plane and $vw$-plane. Top row: Values of $log( \osjml)$ at $z$ = 0 in the $uw$-plane with fixed $ J_\varphi $ = 1759 $\Kpc~\kmsec$, for the various vertical resonances relevant in the solar neighbourhood (the $l$ = 0 resonances are treated in detail in Sect.~3.3). They all appear at relatively large values of $w$ and are very concentrated in $w$, varying very quickly in $u$ as a function of $w$. Bottom row: Values of $ log(\osjml)$ in the  $vw$-plane with fixed $u$ = 0 $\kmsec$. The pattern speed $\Omp$ is that of our fiducial central bar fixed at 1.89 $\Omega_0$.}
            \label{resonanceVelocitySpace}
        \end{figure*}
        
            \begin{figure*}
        \centering
                \includegraphics[scale=0.28]{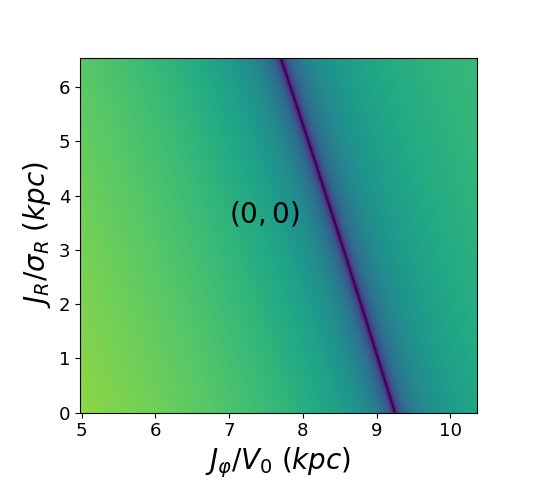}
                \includegraphics[scale=0.28]{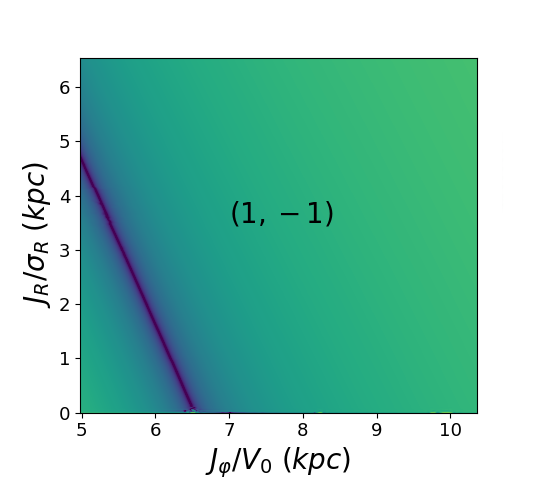}
                \includegraphics[scale=0.28]{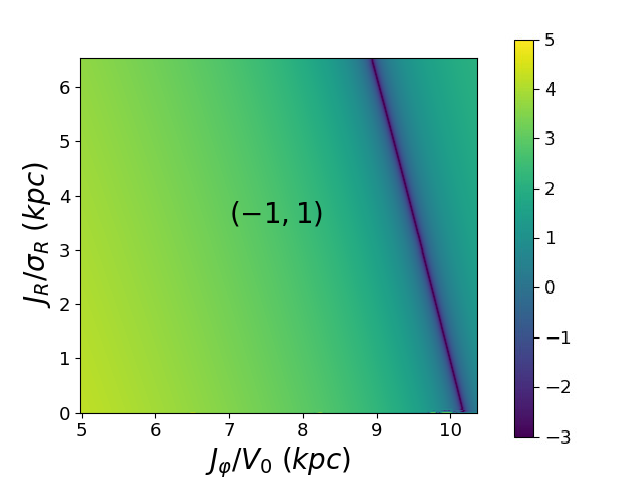}
                \caption{Same as Fig.~\ref{resonanceBarJphi}, but with  some combinations of indices giving rise to resonant zones for $\Omp = 0.84 \, \Omega_0$.}
                \label{resonanceSpiralJphi}
        \end{figure*}
        
            \begin{figure*}
        \centering
                \includegraphics[scale=0.28]{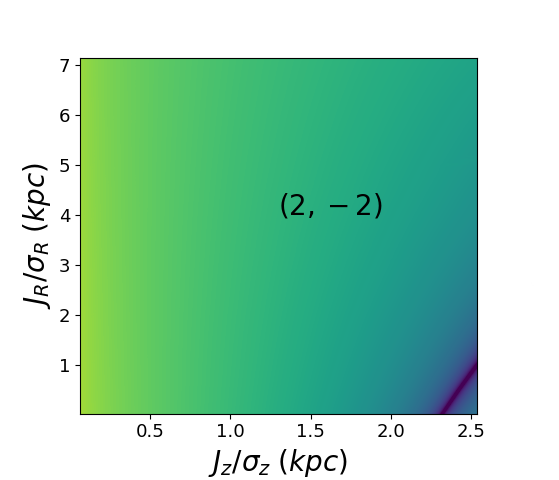}
                \includegraphics[scale=0.28]{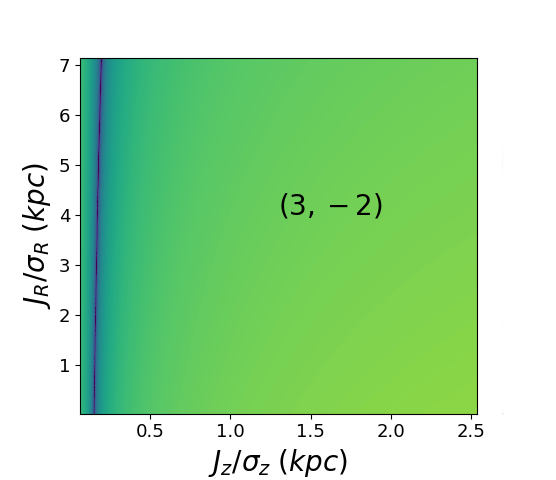}
                \includegraphics[scale=0.28]{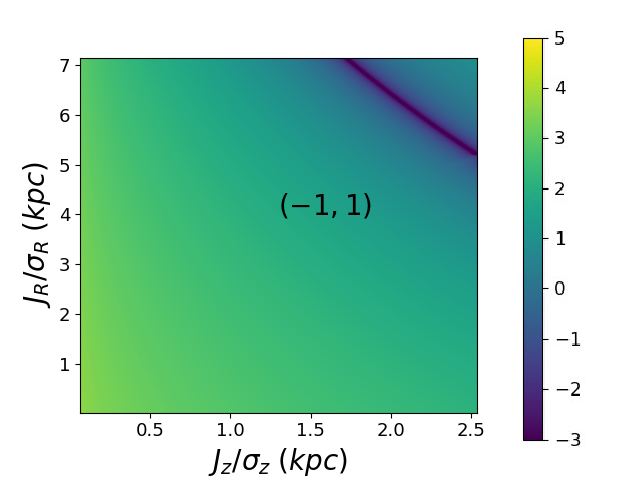}
                \caption{Same as Fig.~\ref{resonanceBarJz}, but with some combinations of indices giving rise to resonant zones for $\Omp = 0.84 \, \Omega_0$.}
                \label{resonanceSpiralJz}
        \end{figure*}

The potential we use for the spiral arms is the following \citep{CoxGomez2002,Monari2016}
        \begin{equation}
                \Phi_{1,\mathrm{sp}}(R,z,\varphi,t) = \Rep \, \left\lbrace \Phi_{\mathrm{a,sp}}(R,z) \, \eexp^{\img m (\varphi - \phis - \Omegasp t)} \right\rbrace,
        \end{equation} where $m=2$, $\Omegasp$ is the pattern speed of the spiral arms, and 
        \begin{equation}
                \Phi_{\mathrm{a,sp}}(R,z) = -\frac{A}{\Rsp KD} \eexp^{\img m \frac{\mathrm{ln}(R/\Rsp)}{{\rm tan}(p)}} \left[ {\rm sech} \left( \frac{Kz}{\beta} \right) \right]^{\beta} ,
        \end{equation} where
    
        \begin{align}
                &K(R) = \frac{2}{R\sin(p)}, \quad \beta(R) = K(R)\hsp[1+0.4K(R)\hsp], \nonumber \\
                &D(R) = \frac{1+K(R)\hsp+0.3[K(R)\hsp]^2}{1+0.3K(R)\hsp}.
        \end{align}  Here $\Rsp=1\,{\rm kpc}$ is the length parameter of the logarithmic spiral potential, $\hsp=0.1\,{\rm kpc}$ the height parameter, $p=-9.9^{\mathrm{o}}$ the pitch angle, $\phis=-26^{\mathrm{o}}$ the phase, and $A=683.4\,{\rm km}^2\,{\rm s}^{-2}$ the amplitude. Hereafter, we  adopt a pattern speed $\Omegasp = 0.84\,\Omega_0$, placing the main resonances away from (or at high azimuthal velocities in) the solar neighbourhood.
        
Within the Galactic plane, the top panel of Fig.~\ref{PotEstSpiral} shows our reconstruction of the spiral potential at the Sun's position, again with 41 complex Fourier coefficients. The accuracy is again below the per cent level as in the bar case, although in the spiral case there is a slight bias towards higher amplitudes with respect to the input spiral potential. This bias is  very small, however,  and does not affect our results. At $z=0.3$~kpc, more complex Fourier coefficients are again needed (Fig.~\ref{PotEstSpiral}, bottom panel), and the accuracy reaches the per cent level, this time without bias.
        
\section{Results and comparison with the epicyclic approximation}

                            \begin{figure*}
                \centering
                \includegraphics[scale=0.092]{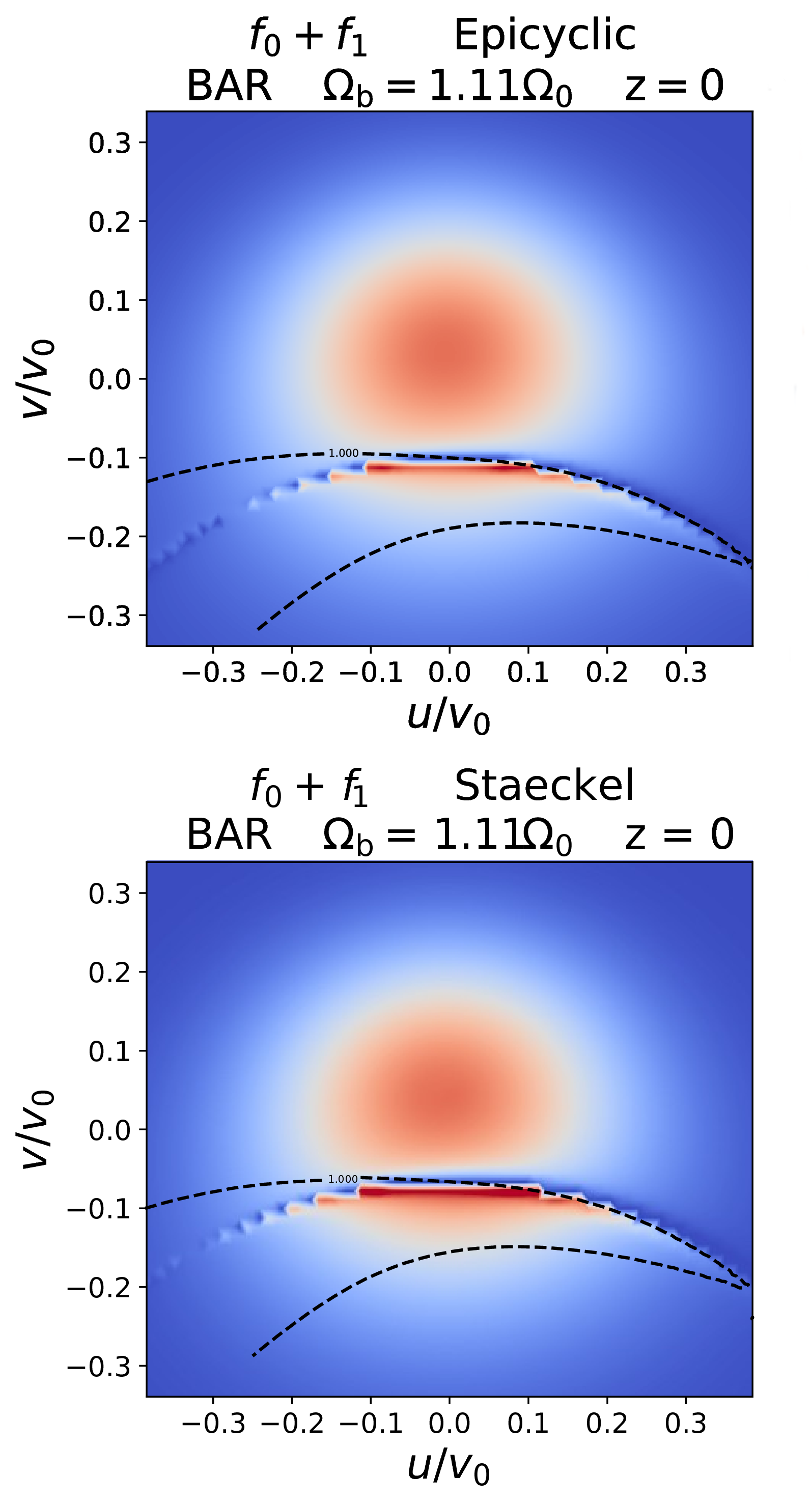}
                \hspace{0.65cm}
                \includegraphics[scale=0.092]{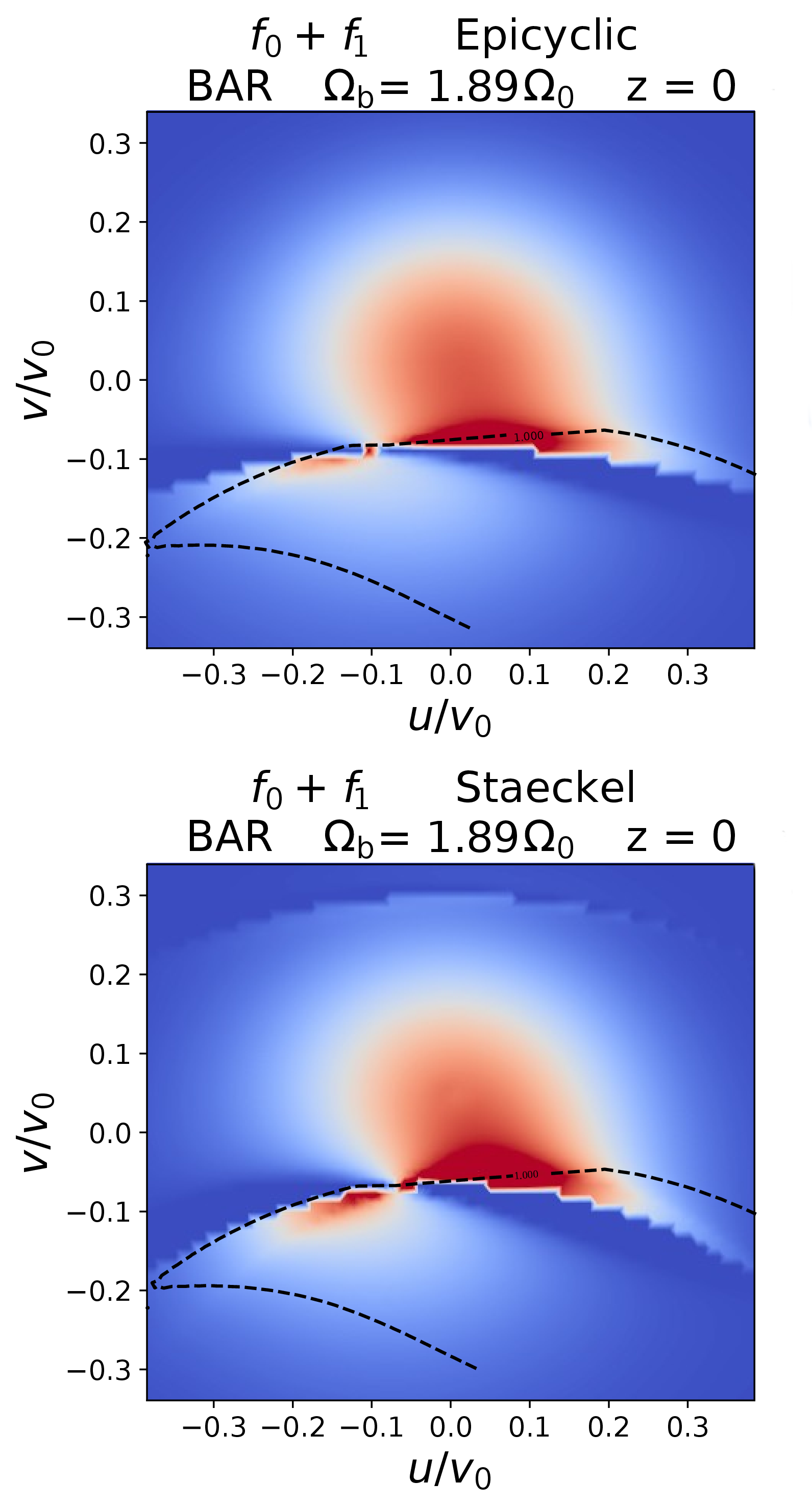}
                \hspace{0.65cm}
                \includegraphics[scale=0.28]{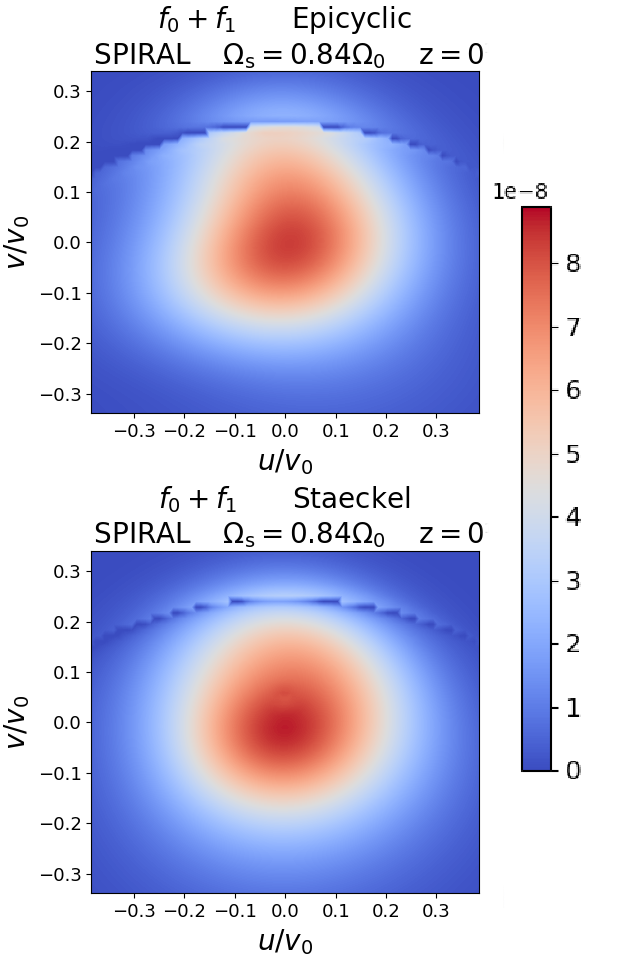}
                \caption{Distribution function from Fig.~\ref{fig_f0} in velocity space at the solar position within the Galactic plane, now perturbed to linear order by a bar (perturbing potential from Sect.~2.4) with pattern speeds $\Omegab=1.16 \, \Omega_0$ (left) and $\Omega_b= 1.89 \Omega_0$ (middle), or by a spiral pattern (perturbing potential from Sect.~2.5) with pattern speed $\Omegasp=0.84 \, \Omega_0$ (right). The black dashed contours represent the zones where k is equal to or less than 1, k being a quantity computed in \citet{Monari2017} that designates the region where the orbits are trapped at the main resonance (the computation used here in the St\"ackel case will be presented in detail in Al Kazwini et al., in preparation). Top row: Epicyclic approximation. Bottom row: St\"ackel fudge. }
                \label{bar_spiral_separate_epivstackel}
        \end{figure*}   
        
                                    \begin{figure*}
                \centering
                \includegraphics[scale=0.35]{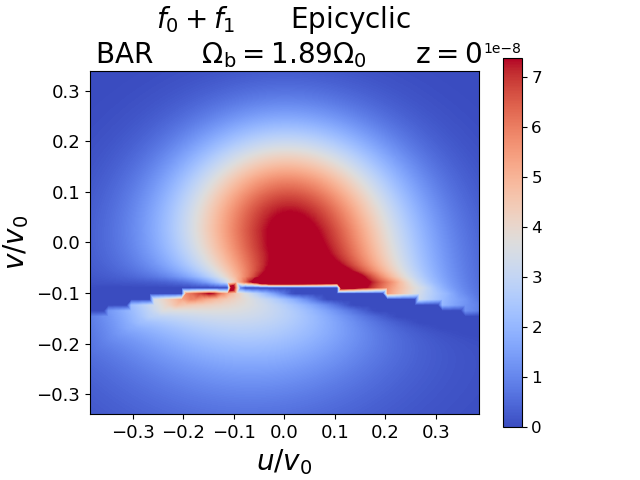}
                \includegraphics[scale=0.35]{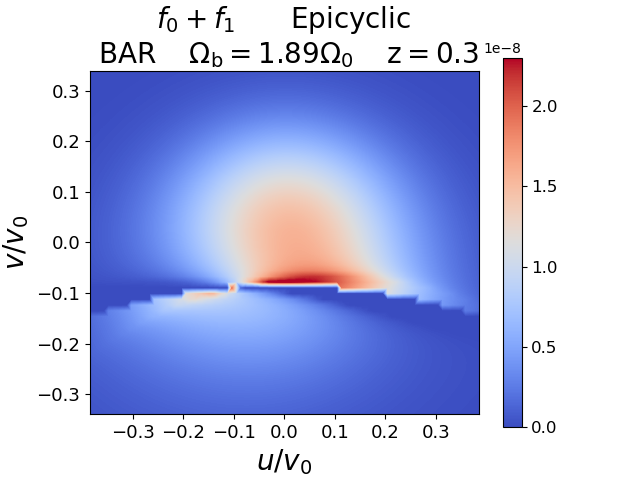}
                \includegraphics[scale=0.35]{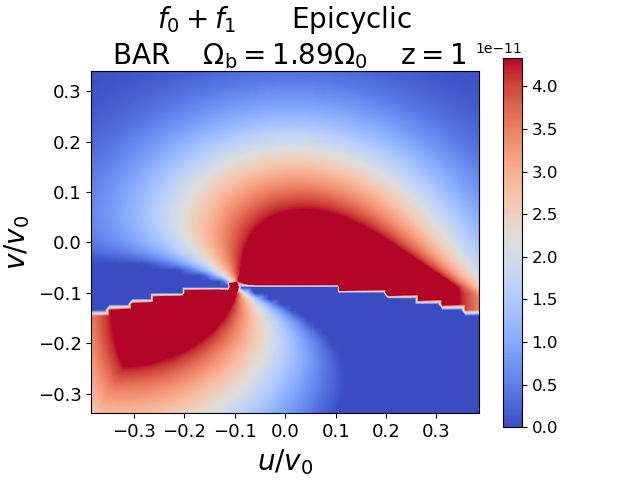}
                \includegraphics[scale=0.35]{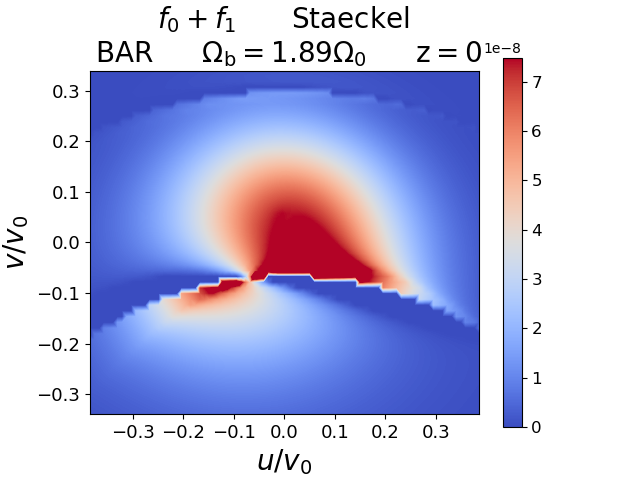}
                \includegraphics[scale=0.35]{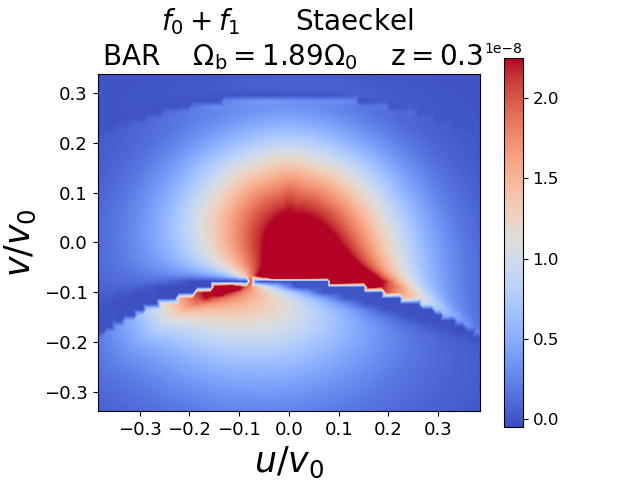}
                \includegraphics[scale=0.35]{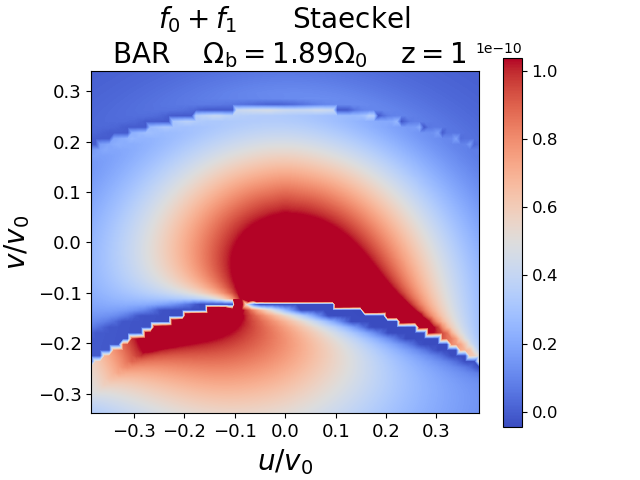}
                \caption{Local stellar velocity distribution perturbed to linear order at the solar galactocentric radius and azimuth at three different heights (left: $z=0 \, \Kpc$, middle: $z=0.3 \, \Kpc$, right: $z=1 \, \Kpc$), when perturbed by a bar (perturbing potential of Sect.~2.4) with pattern speed $\Omega_b= 1.89 \Omega_0$. Top row: Epicyclic approximation. Bottom row: St\"ackel fudge. The scale of the colour bar is different in the upper and lower panels for $z = 1$ kpc.}
                \label{bar_height_epivstackel}
        \end{figure*}
        
        \begin{figure*}
                        \centering
                \includegraphics[scale=0.35]{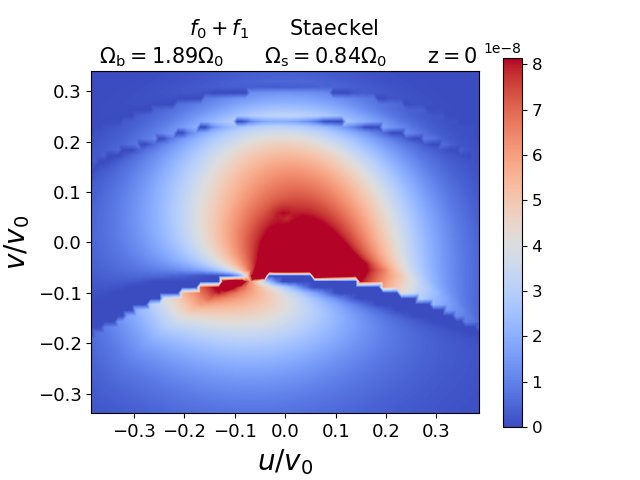}
                \includegraphics[scale=0.35]{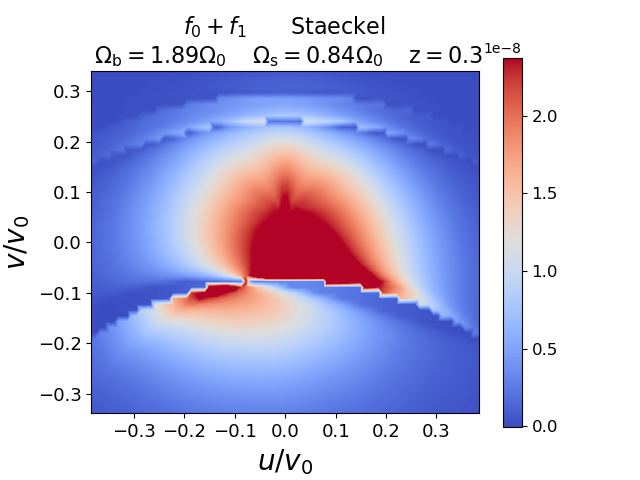}
                \includegraphics[scale=0.35]{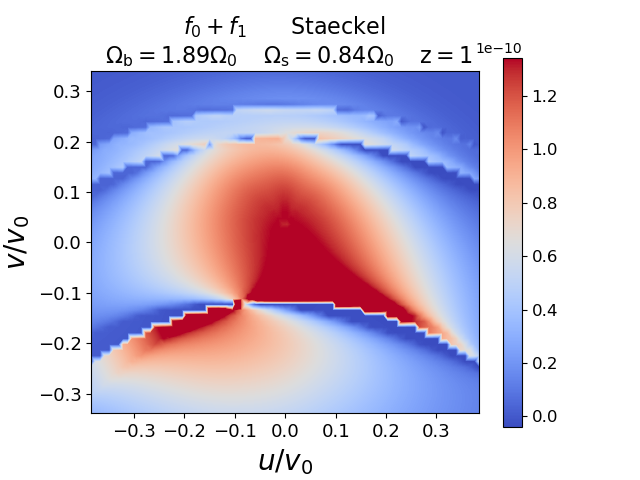}
                \caption{Same as Fig.~\ref{bar_height_epivstackel}, in the St\"ackel fudge case, but now for joint perturbation by a bar (perturbing potential of Sect.~2.4) with pattern speed $\Omega_b= 1.89 \Omega_0$ and a spiral pattern (perturbing potential of Sect.~2.5) with pattern speed $\Omegasp=0.84 \, \Omega_0$.}
                \label{barandspiral_height}
        \end{figure*}

\subsection{Background equilibrium}

From here on we work with a background axisymmetric DF $f_0$ as a sum of two quasi-isothermal DFs \citep{BinneyMcmillan2011} for the thin and thick disc:
\begin{equation}
    f_0(J_R,J_z,J_{\varphi}) = f_{\rm thin} + 0.075 f_{\rm thick}.
    \label{eq:f0}
\end{equation}
The form of each DF is
        \begin{equation}
                f(J_R,J_z,J_{\varphi}) = \frac{\Omega \,\mathrm{exp}(-\Rg/h_R)}{2\,(2\pi)^{3/2}\,\kappa \, \tilde{\sigma}_R^2\,\tilde{\sigma}_z\,z_0} \mathrm{exp} \left( -\frac{J_R\kappa}{\tilde{\sigma}_R^2} - \frac{J_z\nu}{\tilde{\sigma}_z^2} \right)
        ,\end{equation} where $\Rg$, $\Omega$, $\kappa$, and $\nu$ are all functions of $J_\varphi$, and
        \begin{align}
                &\tilde{\sigma}_R(\Rg) = \tilde{\sigma}_R(R_0)\,\mathrm{exp} \left( -\frac{\Rg-R_0}{h_{\sigma_R}} \right), \nonumber \\
                &\tilde{\sigma}_z(\Rg) = \tilde{\sigma}_z(R_0)\,\mathrm{exp} \left( -\frac{\Rg-R_0}{h_{\sigma_z}} \right).
        \end{align} 
        
        For the thin disc DF $f_{\rm thin}$, we choose $h_R=2\,\Kpc$, $z_0=0.3\,\Kpc$, $h_{\sigma_R} = h_{\sigma_z}=10\,\Kpc$, $\tilde{\sigma}_R(R_0)=35\,\kmsec$, and $\tilde{\sigma}_z(R_0)=15\,\kmsec$. For the thick disc DF $f_{\rm thick}$, we choose $h_R=2\,\Kpc$, $z_0=1\,\Kpc$, $h_{\sigma_R}=10\,\Kpc$, $h_{\sigma_z}=5\,\Kpc$, $\tilde{\sigma}_R(R_0)=50\,\kmsec$, and $\tilde{\sigma}_z(R_0)=50\,\kmsec$. Since we normalize the central surface densities of the thin and thick disc to 1, our densities can be multiplied by the appropriate surface density of the relevant stellar population to obtain physical units.
        The  background axisymmetric potential is chosen to be Model I from \citet{BT08}, in which the above equilibrium DF $f_0$ is a good representation of the thin and thick disc components. In this model one has $R_0=8\,\Kpc$ and $v_0= 220 \, \kmsec$.
        
 The top panels of Fig.~\ref{fig_f0} display the $(u,v)$-plane in the solar neighbourhood within the $z=0$ plane (and for $w = - v_z = 0$) for this $f_0$ axisymmetric background, where $u=-v_R$ and $v=v_\varphi - v_0$, obtained by converting velocity-space into action-space through the epicyclic approximation and the St\"ackel fudge from {\tt AGAMA}. The velocity distributions are quite similar. However, as can be seen in the bottom panels of Fig.~\ref{fig_f0}, the epicyclic approximation quickly becomes imprecise outside of the plane as it implies a sharper falloff of the density compared to the better St\"ackel action estimates.
        
\subsection{Resonant zones}

In the case of a perturbation with quasi-static amplitude that has reached its plateau, once the Fourier coefficients representing the perturbing potential have been computed (from the epicyclic approximation or from Eq.~\ref{numFC}) the expression for the perturbed DF can be simply expressed away from resonances with Eq.~\ref{f1M16} as
        \begin{equation}
                f_1(\bJ,\bth,t) = \Rep \, \biggl\lbrace \sum_{j,l=-n}^n f_{jml}\,\eexp^{\img[j\theta_R+m(\theta_{\varphi}-\Omp t)+l\theta_z]} \, \biggr\rbrace ,
        \end{equation} with $n$ the order of the Fourier series (in this paper, $m=2$ in both the bar and spiral cases), and
    \begin{equation}
        f_{jml} = \phi_{jml} \times \frac{j\frac{\partial f_0}{\partial J_R}+m\frac{\partial f_0}{\partial J_\varphi}+l\frac{\partial f_0}{\partial J_z}}{j\omega_R+m(\omega_\varphi-\Omp)+l\omega_z},
    \end{equation}
where $\omega_R$, $\omega_\varphi$, and $\omega_z$ can be approximated as epicyclic frequencies in the epicyclic case or can be determined with {\tt AGAMA}. The denominator of $f_{jml}$ may lead to a divergence in the DF when it approaches zero. Following our notation in Eq.~\ref{eq:omegas},    it can be expressed as
\begin{equation}
    \osjml(J_R,J_\varphi,J_z) = j\omega_R+m(\omega_\varphi-\Omp)+l\omega_z.
\end{equation}
The amount of the resonances is limited in the epicyclic case because, by construction, indices run only over the values $j= \{-1, 0, 1\}$ and $l=\{-2, 0, 2\}$, but they can be much more numerous in the more accurate {\tt AGAMA} case. For the bar potential of Eq.~\ref{barpot} and Eq.~\ref{barpot2}, and choosing a pattern speed $\Omp = 1.89 \Omega_0$ as for our fiducial bar model, we explore in Fig.~\ref{resonanceBarJphi} and Fig.~\ref{resonanceBarJz}, the values of $\osjml(J_R,J_\varphi,J_z)$ in action space when varying the pair of integer indices ($j$, $l$). The actions are renormalized by the radial velocity dispersion of the thin disc, circular velocity, and vertical velocity dispersion of the thin disc at the Sun, respectively, to only display a relevant range of actions. Exploring indices in the range $[-4, +4]$, it is clear that most combinations do not induce a resonance that is relevant to the dynamics of the solar neighbourhood. We only display in Fig.~\ref{resonanceBarJphi} and Fig.~\ref{resonanceBarJz} the combination of indices (in addition to the corotation) for which a resonant zone appears in the plotted region of action space. It is clear that very few low-order resonances are indeed present in the range of actions that are truly relevant for the solar neighbourhood. 

 To date our method  has not been adapted to the projection of the DF on a plane in action space or local velocity space, and therefore works best in 3D. Therefore, we show in Fig.~\ref{resonanceVelocitySpace} some slices in velocity space at $z=0$, denoting the location of the vertical resonances (i.e. resonances involving a non-zero $l$, hence involving the vertical frequency) either for a fixed value of the azimuthal velocity (and action) or for a fixed value of the radial velocity. Identifying such resonances in the $vw$-plane and $uw$-plane should allow  new types of constraints to be put on the pattern speed of internal perturbers and the vertical shape of the potential of the Galaxy.
 
 Interestingly, most of these resonances are very concentrated in $w$ and vary quickly both in $u$ and $v$ as a function of $w$, making them elusive to find when stacking tracer stars in any 2D plane of velocity space, but in principle they stand out in thin slices of velocity space. Concretely, when considering a change of 10~$\kmsec$ in vertical velocity from 5 to 15~$\kmsec$, the corresponding change in the location of the vertical resonance in $v$ within the $uv$-plane is always larger than 10~$\kmsec$ and typically larger (sometimes much larger) than 30~$\kmsec$. 
 
 Moreover, the signature of these vertical resonances in the $uv$-plane is rather thin, typically of the order of the $\kmsec$, hence much thinner than the displacement of the resonance with $w$. This means that, when investigating the $uv$-plane, vertical resonances should mostly be washed out as soon as the investigated slice is thick enough. Therefore, when investigating the DF in the $uv$-plane in the next subsection, we  limit ourselves to the effect of $l=0$ resonances.
 
 As displayed in Fig.~\ref{resonanceSpiralJphi} and Fig.~\ref{resonanceSpiralJz}, for a lower pattern speed $\Omp = 0.84 \, \Omega_0$, corresponding to the pattern speed of our fiducial spiral potential,  a smaller number of  vertical resonances are prominent in the solar neighbourhood.
 
 While a specific treatment is needed in these resonant zones \citep[e.g.][]{Monari2017}, the signature of the resonances (and thus their location in velocity space) can clearly be identified with our linear perturbation method, and the linear perturbation treatment hereafter should accurately describe the deformations of velocity space outside of these resonant zones.

\subsection{Comparing the perturbed DF for different action estimates}

        We are now in a position to compare the linear deformation of local velocity space for different action estimates, namely the epicyclic case used in previous works and the more accurate {\tt AGAMA} action estimates. Since our method works best for now in 3D velocity space, we  limit ourselves to slices of zero vertical velocity at different heights and to $l=0$ resonances.
        
        Figure~\ref{bar_spiral_separate_epivstackel} displays the $f_0+f_1$ linearly perturbed distribution function at the position of the Sun in the Galactic plane for the bar potential of Sect.~2.4 and two different pattern speeds, and for the spiral potential of Sect.~2.5. As in \citet{Monari2017a}, whenever $f_1 > f_0$, we cap $f_1$ at the value of $f_0$ to roughly represent the resonant zone. The more rigorous approach, which we leave to further work in the context of {\tt AGAMA} actions (Al Kazwini et al., in prep.), is to treat the DF with the method of \citet{Monari2017} in these regions. However, while the DF within the resonant zone is not well modelled by the present method, the {\it location} and global shape of resonances should be well reproduced. We indeed highlight in Fig.~\ref{bar_spiral_separate_epivstackel} the zone occupied by trapped orbits at the corotation ($\Omegab=1.16 \, \Omega_0$) and OLR ($\Omega_b= 1.89 \Omega_0$) of the bar, as determined with the method of \citet{Monari2017} both in the epicyclic and {\tt AGAMA} cases (Al Kazwini et al. in prep.). While the {\it quantitative} enhancement of the DF will be slightly different from our linear treatment in these trapping zones, it is clear that the location of the resonant deformation is well captured by the method, as expected.
        The linear deformation outside of the resonant zones should be well described by our method as well. Interestingly enough, the linear deformation due to the bar is generally stronger in the {\tt AGAMA} case, and that due to the spiral is weaker in the {\tt AGAMA} case. This means that reproducing the effect of spiral arms on the local velocity distribution might require a higher amplitude when considering an accurate estimate of the action-angle variables rather than the epicyclic approximation. We speculate that this is related to the inaccuracy of the reconstruction of the potential in the epicyclic case, which causes different biases in the spiral and bar cases.
        
        The case of the pattern speed of the bar being $1.89\,\Omega_0$  would correspond to a configuration where the Hercules stream at negative $u$ and negative $v$ corresponds to the $2:1$ outer Lindblad resonance of the bar \citep[e.g.][]{Dehnen2000,Minchev,Monari2017a,Fragkoudi}. Although this happens in the resonant zone, it is interesting to note that this feature is less squashed in the more realistic {\tt AGAMA} case. Moreover, a resonance unnoticed within the epicyclic approximation appears at high azimuthal velocities: we can identify this resonance as the outer $1:1$ resonance of the bar \citep{Dehnen2000}. In the spiral case, the resonant ridge at large azimuthal velocities can be identified as the corotation of the spiral pattern.
        
        Figure~\ref{bar_height_epivstackel} displays the linear deformation due to the bar, for the case of  pattern speed of $1.89\,\Omega_0$, at different heights above the Galactic plane, both in the epicyclic and {\tt AGAMA} cases. We again restrict ourselves to a zero vertical velocity slice and $l=0$ resonances. As can be seen in this figure, the epicyclic approximation quickly becomes imprecise at large heights because it implies a stronger falloff of the density with height (as already noted in Fig.~\ref{fig_f0}) while not changing the azimuthal velocity distribution (and the location of resonances in $v$) due to the hypothesis of complete decoupling of vertical motions. 
        
        In the {\tt AGAMA} case the azimuthal velocity distribution is affected by a larger asymmetric drift at large heights, and the location of the outer Lindblad resonance of the bar in the $uv$-plane is also displaced to lower azimuthal $v$ at larger heights. This occurs because at fixed $J_\varphi$ the azimuthal and radial frequencies computed with {\tt AGAMA} are lower at higher $z$, meaning that one needs to reach lower $J_\varphi$ (corresponding to orbits whose guiding radii are in the inner Galaxy) to reach the resonance. 
        
        This trend is most clearly visible at $z=1$~kpc, where the epicyclic approximation does not accurately represent  the location of the Hercules feature compared to the {\tt AGAMA} case. Interestingly, comparing the displacement with height of the OLR in the case of a bar with pattern speed $1.89 \, \Omega_0$ with that of the corotation in the case of a $1.16 \, \Omega_0$ pattern speed, we noted that the corotation location in the $uv$-plane is more displaced than the OLR. This is because the corotation only depends on the azimuthal frequency, while the OLR depends on a combination of the azimuthal and radial frequencies. 
        On the other hand, with the presently assumed background potential, we found that the displacement with height was rather independent of the pattern speed and therefore of the location of the resonance in local velocity space. We found a gradient in $v$ of 8~${\rm km} {\rm s}^{-1} {\rm kpc}^{-1}$ for the corotation, 6~${\rm km} {\rm s}^{-1} {\rm kpc}^{-1}$ for the OLR, and 4~${\rm km} {\rm s}^{-1} {\rm kpc}^{-1}$ for the $1:1$ resonance. This different displacement can also be seen when linearly adding the effect of the bar and spiral in Fig.~\ref{barandspiral_height}, where the spacing between the $1:1$ resonance of the bar and that of the corotation of the spiral increases with height.
        
        Quantitatively,  these displacements  depend strongly on the background Galactic potential. This means that once  the resonances potentially responsible for moving groups in the solar neighbourhood have been identified, studying their position in the $uv$-plane as a function of $z$ can in principle be a powerful new way to constrain the 3D structure of the Galactic potential. This cannot be done within the epicyclic approximation. We note that marginalizing over vertical velocities instead of taking a zero-velocity slice would not compensate for these variations of the location of resonances with height but would only enhance the effect. In practice, we investigated the displacement of the location of the in-plane OLR  with vertical velocities. For $w=50 \, \kmsec$ the displacement compared to $w=0 \, \kmsec$ in terms of the $v$-location of the resonance at $z=1$~kpc is 8~$\kmsec$ , always towards lower azimuthal velocities; however, the signal will  always be dominated by the lowest $w$ values due to the vertical orbital structure of the disc.

        \section{Adding the temporal evolution}
        
        In the previous sections we always consider a constant amplitude for the perturbing potential in order to determine an analytical expression for the perturbed DF. In this section we investigate the time dependence of the DF by choosing a time-varying amplitude for the perturbing potential.
        
        \subsection{Time-varying amplitude function}
        
        The expression we  use for the  time-dependent function $g$ controlling the amplitude of the perturbation during its growth is
        \begin{align}
                g(t) = \frac{1-\cos\left(\pi t/\tf \right)}{2} ,
        \end{align} 
        where $\tf$ is the time at which the perturbation is completely formed, expressed in Gyr. We  consider $\tf = 0.5\,\Gyr$.
        
        The motivation for this choice of growth function is its analytic simplicity, having a function starting from exactly zero at the origin, and smooth over the whole considered range. The first derivative, $[\pi/(2\tf)]\sin(\pi t / \tf)$, assures the continuity at $0$ and $\tf$ with both stages, fixed at $0$ for $t\leq 0$ and at $1$ for $t\geq \tf$ (the first derivative is thus equal to $0$ at $0$ and $\tf$).
        
        \subsection{Time-dependent perturbed distribution function}
        
        We now take the integral of Eq.~(\ref{pert_df_2}), restricted to $\left[0,t \right]$ (because the $g$ function is equal to $0$ on $ \left]-\infty,0\right]$) and integrate by parts.  We take $\phi_{\bn}(\bJ', t') = g(t') \, h(t') \, \phi_{\bn}(\bJ)$, with $h(t') = \eexp^{-\img m \Omp t'}$, and we define
        \begin{equation}
            \eta(t) \equiv \frac{\eexp^{\img \otn(t)}}{\img \osn}
            \rightarrow
                \deriv \eta =\eexp^{\img \otn(t)}\deriv t, 
        \end{equation}
        allowing us to rewrite Eq.~(\ref{pert_df_2}) as
        \begin{align} 
        f_1(\bJ,\bth,t) = \Rep \, \biggl\lbrace \, &\img \frac{\partial f_0}{\partial \bJ}(\bJ) \cdot \sum_{\bn} \bn \phi_{\bn}(\bJ)
        \int_{0}^{t} g(t') \frac{\de\eta}{\de t'}(t') \, \deriv t' \biggr\rbrace.
\end{align}
        
    \noindent We can now integrate by parts
        \begin{equation}\label{eq:parts}
                \int_0^t g(t')\frac{\de\eta}{\de t'}(t') \, \deriv t' = 
                \left[
                        g(t') \eta(t')
                \right]_0^t
                - \int_0^t \frac{\deriv g}{\deriv t'}(t') \, \eta(t') \, \deriv t',
        \end{equation}

        \noindent and since $g(0)=0$,
        \begin{equation}\label{eq:geta}
                \left[
                        g(t') \eta(t')
                \right]_0^t =
                g(t) \eta(t).
        \end{equation}

    To calculate the second part of the integral, since $\deriv g(t)/\deriv t = \pi/(2\tf) \sin(\pi t/\tf)$, we write,
    \begin{equation}\label{eq:integrand}
                \int_0^t \frac{\deriv g}{\deriv t'}(t') \eta(t') \deriv t' =
                \frac{\pi}{2\tf}\frac{1}{\img \osn}
                \int_0^t \sin\left(\frac{\pi t'}{\tf}\right)\eexp^{\img \otn(t')}\deriv t'.
    \end{equation}
        We look for a primitive $G$ of $\sin(\pi t/\tf)\eexp^{\img \otn(t)}$ of the form 
        \begin{equation}\label{eq:G}
            G(t) = \left[ A \cos\left(\frac{\pi t}{\tf}\right)+B \sin\left(\frac{\pi t}{\tf}\right) \right]\eexp^{\img \otn(t)}.
        \end{equation}
        Deriving $G(t)$ with respect to $t$, and equating it to the integrand in Eq.~\ref{eq:integrand} we get
        \begin{equation}
                B\frac{\pi}{\tf} + A\,\img \osn = 0 \; \; \; \;
                \mathrm{and} \; \; \; \;
                B\,\img \osn - A\frac{\pi}{\tf} = 1,
        \end{equation}
        which leads to
        \begin{equation}
                A = \frac{\pi/\tf}{\osn^2-(\pi/\tf)^2} \; \; \; \; 
                \mathrm{and} 
                \; \; \; \;
                B = \frac{-\img \osn}{\osn^2-(\pi/\tf)^2}.
        \end{equation}
        Substituting the Eqs.~(\ref{eq:geta}) and (\ref{eq:G}) into Eq.~(\ref{eq:parts}) results in the following expression for the perturbed DF (taking the real part of the expression)
        \begin{equation}
        \begin{split}
              & f_1\left(\bJ,\bth,t\right)  = 
                \frac{\partial f_0}{\partial\bJ}(\bJ) \cdot
                \sum_{\bn} \bn\,\phi_{\bn}(\bJ) \, \times & \\ &  \Bigg[ \,
                \frac{1}{2} \left( 1-\cos\left(\frac{\pi t}{\tf}\right)\right) \frac{\eexp^{\img \otn}}{\osn} 
                - \frac{\pi}{2\tf} \, \frac{1}{\osn} 
                \frac{1}{\osn^2-(\pi/\tf)^2} \, \times & \\ & 
                        \left( \left( \frac{\pi}{\tf} \cos\left(\frac{\pi t}{\tf}\right) - \img \osn \sin\left(\frac{\pi t}{\tf}\right) \right) \,\eexp^{\img \otn} - \frac{\pi}{\tf}\,\eexp^{\img (\otn - \osn t)} \right)
                \, \Bigg].
                \end{split}
        \end{equation}

It should be noted that we do not exactly recover the static case at $t=t_f$ because not {\it all} derivatives of $g(t)$ are strictly zero at the initial and final time, as assumed in M16. If  a true plateau is reached after $t_f$ in an analytic fashion, the function would nevertheless converge towards the static case. How quickly this would happen is not trivial to compute. We can however compute an upper limit based on the formalism of \citet{Monari2017}. Considering that the most trapped orbits have their slow variables following the behaviour of a harmonic oscillator, and taking $2 \pi$ over the frequency of this harmonic oscillator as a characteristic time for phase-mixing, we obtain a characteristic time of the order of 2~Gyr.

\begin{figure*}
                \centering
                \includegraphics[scale=0.50]{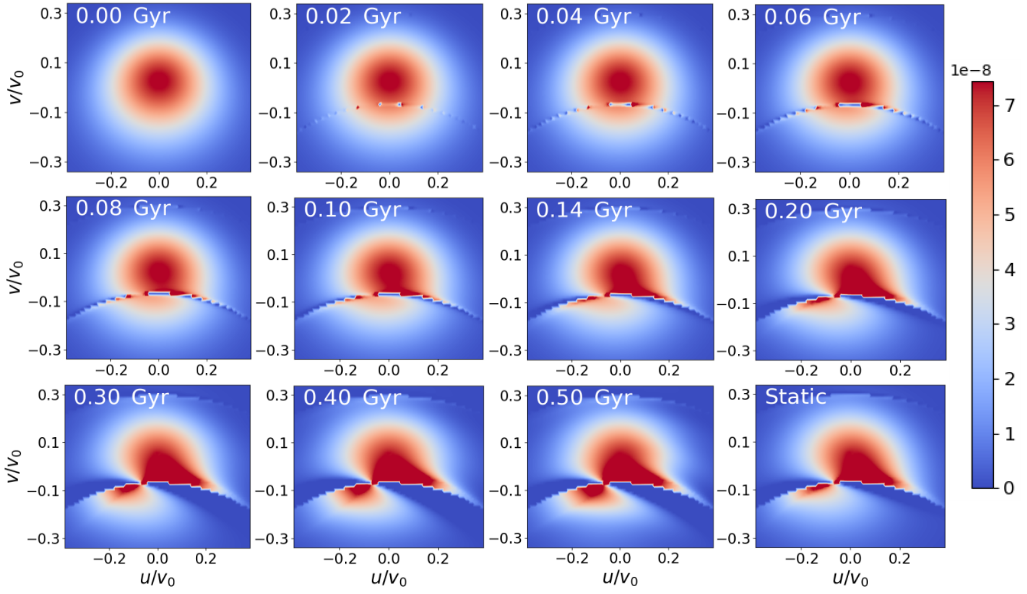}
                \caption{Local stellar velocity distribution perturbed to linear order in the Galactic plane by the bar of Sect.~2.4 with $\Omegab = 1.89\Omega_0$ with the St\"ackel fudge, and an amplitude of the bar growing as described in Sect.~4.1. The first 11 panels display the temporal evolution of the perturbation. The last panel displays the stationary case. The amplitude of the bar goes from $0$ at $t=0$ to its plateau ($g(t)=1$) at $t=0.5\Gyr$.} 
                \label{fastbartimevo}
        \end{figure*}

Now we can study analytically how the linear response to a fiducial bar with $\Omegab = 1.89\Omega_0$ evolves with time. As before the method is not strictly valid at resonances, where a treatment like that used in \citet{Monari2017} must be applied \citep[see also][]{Binney2020a, Binney2020b}.  It is nevertheless interesting to see in Fig.~\ref{fastbartimevo} how the linear deformation of the velocity plane evolves with time near resonances (in a patch co-moving with the bar, hence at a constant azimuthal angle to the bar), while the amplitude of the perturbation grows. The effect of the OLR appears as soon as the perturbation starts to grow. As it progressively grows, the two linear modes in the DF separate and lead to a velocity plane already very much resembling the stationary form of the perturbed DF after $0.25\Gyr$, that is when $g(t)=0.5$ and the perturbation is half-formed. In the absence of a pattern speed variation, it is therefore not necessarily obvious to disentangle the effect of a bar whose amplitude is growing from that of a fully formed bar with larger and constant amplitude.
        
\section{Conclusion}

Starting from the formalism exposed in M16, we proposed a more accurate way to determine the DF of the Galactic disc perturbed to linear order by a non-axisymmetric perturbation, using a more accurate action-angle coordinate system. First, we used the torus mapping from {\tt AGAMA} to numerically compute the perturbing potential in action-angle coordinates as a Fourier series expansion over the angles. We showed that we could estimate typical non-axisymmetric perturbing potentials with an accuracy at the per cent level. The algorithm can be applied to any perturbing potential, including non-plane symmetric vertical perturbations, which will be particularly important when studying the vertical perturbations of the disc with similar methods (Rozier et al., in prep.).

We then computed the DF perturbed to linear order by a typical bar or spiral potential (or a linear combination of both), and computed the local stellar velocity distribution by converting velocities to actions and angles through the St\"ackel fudge implemented in {\tt AGAMA}. The results were compared to those obtained by using the epicyclic approximation. The linear deformation due to the bar is generally stronger in the {\tt AGAMA} case,  and  that  due  to  the  spiral  is  weaker  in  the {\tt AGAMA} case. This means that reproducing the effect of spiral arms on the local velocity distribution might require a higher amplitude when considering  an  accurate  estimate  of  the  action-angle  variables rather than the epicyclic approximation. Most importantly, the epicyclic approximation is inadequate at large heights and does not change the azimuthal velocity location of the resonances due to the hypothesis of complete decoupling of vertical motions. In the {\tt AGAMA} case instead, the locations of resonances are displaced to lower azimuthal $v$ at larger heights. With the background potential used in this paper, we found a displacement in $v$ of 8~${\rm km} {\rm s}^{-1} {\rm kpc}^{-1}$ for the corotation, 6~${\rm km} {\rm s}^{-1} {\rm kpc}^{-1}$ for the OLR and 4~${\rm km} {\rm s}^{-1} {\rm kpc}^{-1}$ for the $1:1$ resonance. Thus, the position of moving groups in the $uv$-plane as a function of $z$ can be a powerful way to constrain the 3D structure of the Galactic potential. The key to exploring this will be the DR3 of Gaia \citep{Brown2019} with its $\sim 35$~million radial velocities allowing us to better probe the $z$-axis above and below the Milky Way plane.

Finally, the temporal treatment is also an improvement over M16. We applied it to the case of a bar of growing amplitude, with an analytic evolution of the amplitude. As the bar progressively grows, the two linear modes in the DF separate, and lead to a velocity plane already very much resembling the stationary form of the perturbed DF once the perturbation is half-formed. In the absence of a pattern speed variation, it is therefore not necessarily obvious to disentangle the effect of a bar whose amplitude is growing from that of a fully formed bar with larger and constant amplitude. We  explored here a peculiar form of the growth function motivated by its analytic simplicity. If the perturbation grows by linear instability, exponential growth will be more realistic. Numerical experiments are usually well fitted by a logistic function (exponential growth at the beginning and saturation to the plateau). One problem for our treatment is that the logistic function is never strictly equal to 0. In addition, there is hope that similar analytical simplifications such as those for the amplitude growth studied here can also be made with this function, which we will investigate in the future.

While the form of the DF is not well estimated in the resonant zones with the linear perturbations presented in this paper, the signature of the resonances (and thus their location in velocity space) can clearly be identified with this linear perturbation method. The more rigorous approach, which we leave to further works in the context of {\tt AGAMA} actions (Al Kazwini et al., in prep.), is to treat the DF with a method like that of \citet{Monari2017} in these regions, patching these results over the linear deformation computed here. Another caveat is that the torus mapping was used to express the perturbing potential in actions and angles, but for the estimate of the local stellar velocity field, we made use of the less precise St\"ackel fudge method. Therefore, another promising way for improvement would be to use the new {\tt ACTIONFINDER} deep-learning algorithm \citep{actionfinder} to make the reverse transformation. Finally, the results presented in this paper were   obtained in 3D action and velocity spaces, and were mostly presented in 2D slices: it would therefore be particularly useful to improve our algorithm by including a marginalization over any axis, for instance marginalizing over vertical velocities. This is  computationally more intensive but should not, {a priori},   pose any conceptual problem.

The tools presented in this paper will be useful for a thorough analytical dynamical modelling of the complex velocity distribution of Milky Way disc stars as measured by past and upcoming Gaia data releases. These tools will also be useful for fully self-consistent treatments of the response of the disc to external perturbations. The ultimate goal is  to adjust models to the exquisite data from Gaia, which cannot be done properly with N-body simulations due to the vast parameter space to explore. The theoretical tools and the new code presented in this paper consequently represent a useful step in this direction.

\begin{acknowledgements}
We thank the referee, Dr Paul McMillan, for a thoughtful and constructive report which has helped improving the paper and correcting a small bug in the first version. AS, BF, GM, SR, PR and RI acknowledge funding from the Agence Nationale de la Recherche (ANR project ANR-18-CE31-0006 and ANR-19-CE31-0017) and from the European Research Council (ERC) under the European Union's Horizon 2020 research and innovation programme (grant agreement No. 834148).
\end{acknowledgements}

\bibliographystyle{aa}
\bibliography{f1timedep}

\begin{appendix}

        \section{PERDIGAL code presentation}
        
\begin{figure*}
                \centering
                \includegraphics[scale=0.40]{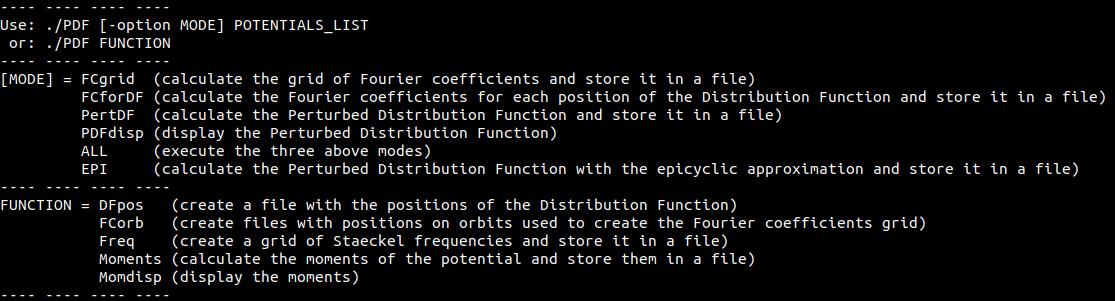}
                \caption{Guide for the use of the code. The first three modes deal with the determination of the Fourier coefficients when expanding in series the perturbing potential (FCgrid and FCforDF), and the calculation of the perturbed distribution function (PertDF). PDFdisp allows this function to be displayed as a distribution of velocities. ALL regroups these   modes into one, but it is not recommended because of the need to check the Fourier decomposition. EPI directly calculates  the perturbed distribution function under the epicyclic approximation, as explained in M16. The FUNCTIONs correspond to different operations required before using the MODEs. DFpos creates the file storing the positions at which the distribution function will be determined. FCorb creates several files containing positions from the many orbits needed to calculate the Fourier coefficients in the FCgrid step. } 
                \label{UsageProg}
        \end{figure*}

        All the results shown in this article were obtained with a single code, written in C++ and Python, that calculates both the perturbing potential in action-angle coordinates and the perturbed DF. The code is named  the PERturbed DIstribution functions for the GALactic disc ({\tt
PERDIGAL}) and will be made available on request, although it may eventually be embedded in a larger Galactic dynamics toolkit. Launching the code without argument gives the explanations shown in Fig.~\ref{UsageProg}.
  
    \paragraph{}
    The calculation of the perturbed DF consists of five steps: the creation of a file storing the positions of the DF (DFpos), the creation (via {\tt AGAMA}) of several files containing positions from the many  orbits (FCorb) required for the following step, the determination of Fourier coefficients for the perturbing potential to create the grid (FCgrid), the determination of the Fourier coefficients for each position which the DF will be calculated at (FCforDF), and finally the calculation of the perturbed DF (PertDF).The mode ALL processes the three  steps (FCgrid, FCforDF, and PertDF) at one time, without saving
    Fourier coeffcients from FCgrid and FCforDF in files. However, ALL should not be used without certainty about the decomposition of the potential. This process strongly depends on  the initial conditions fed to the code, and a check after each part of the calculation is recommended. Finally, PDFdisp displays the perturbed DF using Matplotlib.
        
\begin{figure*}
                \centering
                \includegraphics[scale=0.32]{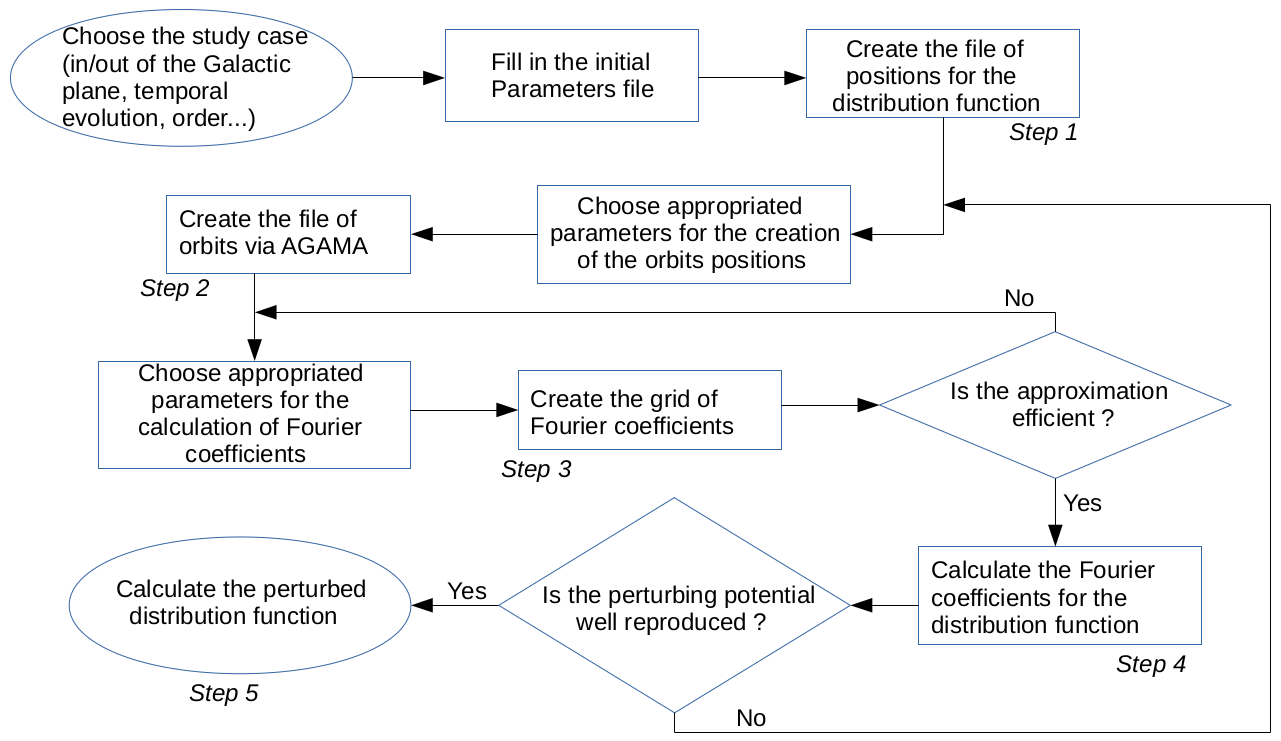}
                \caption{Logigram recalling the procedure of the calculation of the perturbed DF. The different steps 1-5 correspond  respectively to the modes DFpos, FCorb, FCgrid, FCforDF, and PertDF. There are two cases where the result of the code needs to be verified: after the third step (to check that the decomposition
is correct) and after the fourth step (to check that the perturbing
potential is well reproduced).} 
                \label{Logi}
        \end{figure*}

        \paragraph{}
        All useful parameters are stored in two particular files and can be modified without compiling the code. They contain parameters for the calculation of Fourier coefficients, parameters for the perturbing potentials, and others depending on  whether the DF is determined at first or second order, or with a time-dependent amplitude for the perturbing potential. Figure \ref{Logi} is a logigram recalling the procedure of the calculation of the perturbed DF.
        
\end{appendix}



\end{document}